\documentclass[10pt,preprintnumbers,amsmath,amssymb,showpacs]{revtex4}
\usepackage{graphics}
\usepackage{amsmath}
\usepackage{epsfig}
\usepackage{epsf}
\usepackage{graphicx}
\begin{document}
\title{Extraction of the charged pion
polarizabilities from radiative charged pion photoproduction in Heavy Baryon Chiral Perturbation Theory}
\author{Chung Wen Kao$^{1}$, Blaine E. Norum$^2$, and Kebin
Wang$^2$}
\address {$^1$Department of
Physics, Chung Yuan Christian University, Chung-Li 32023, Taiwan}
\address{$^2$Department of Physics, University of Virginia,
Charlottesville, VA 22904-4714, USA}
\date{\today}
\begin{abstract}
We analyze the
amplitude of radiative charged pion photoproduction within the framework of
heavy baryon chiral perturbation theory (HBChPT) and discuss the best experimental
setup for the extraction of the charged pion polarizabilities from the differential
cross section. We find that the contributions from two unknown low energy constants(LECs)
in the $\pi N$ chiral Lagrangian at order $p^{3}$
are comparable with the contributions of the charged pion polarizabilities.
As a result, it is necessary to take these two LECs' effects into account.
Furthermore,
we discuss the applicability of the extrapolation method and conclude that this method
is applicable only if the polarization vector of the incoming photon is perpendicular
to the scattering plane in the center of mass frame of the final $\gamma-\pi$ system.

\end{abstract}
\maketitle
\bigskip
\section{Introduction}
Electric ($\alpha$) and magnetic ($\beta$) polarizabilities
characterize the global responses of a composite system to external
electric and magnetic fields. They provide precious
information about the inner structure of the composite system.
Since the pion is the simplest composite system bound by the strong interaction, its
polarizabilities are fundamental benchmark of QCD in the realm of confinement,
and an accurate determination of the charged pion polarizabilities is highly desirable.
The charged pion polarizabilities have been calculated by chiral perturbation theory(ChPT).
The predictions of ChPT at ${\cal O}(p^4)$
read \cite{BF88}:
\begin{equation}\alpha_{\pi^{\pm}}=-\beta_{\pi^{\pm}}=
\frac{e^{2}}{4\pi}\cdot\frac{2}{m_{\pi}(4\pi F_{\pi})^{2}}
\cdot\frac{\bar{l}_{6}-\bar{l}_{5}}{6}=\frac{\alpha_{em}h_{A}}
{\sqrt{2}F_{\pi}m_{\pi}}=(2.64\pm 0.09 )\times 10^{-4}fm^{3},
\label{ChPT}
\end{equation}
where $\bar{l}_{6}-\bar{l}_{5}$ is a linear combination of
parameters of the Gasser and Leutwyler Lagrangian \cite{Gasser};
$F_{\pi}=93.1\ MeV$ is the pion decay constant; and
$h_{A}=(0.0115\pm 0.0004)/m_{\pi}$ \cite{axel} is the axial vector coupling
constant.
The next-to-leading-order result has been calculated at ${\cal O}(p^{6})$ in
ChPT \cite{BGS94,B97}, and changes the result shown in (\ref{ChPT}) very
little:
\begin{equation}(\alpha+\beta)_{\pi^{\pm}}=(0.3\pm 0.1)\times
10^{-4}fm^{3},\,\,\,(\alpha-\beta)_{\pi^{\pm}}=(4.4\pm 1.0)\times 10^{-4}fm^{3}.
\label{p6}
\end{equation}
Therefore, the measurement of the charged pion polarizabilities becomes an excellent
test of chiral dynamics. There is also a preliminary lattice result by using
background field technique which gives $\alpha_{\pi^{\pm}}=(3.4\pm 0.4)\times 10^{-4}fm^{3}$
\cite{lattice}.
\newline
\indent
Usually the polarizabilities of hadrons are extracted through
Compton scattering.
When the Compton scattering amplitudes are expanded in the energy of the final photon,
its leading order terms
are given by the Thomson limit which only depends
on the charge and the mass of the target.
Genuine structure effects first appear at second order and are parametrized
in terms of polarizabilities:
\begin{equation}T_{\gamma\pi\rightarrow\gamma\pi}=
T_{B}+4\pi\omega\omega'[(\vec{\epsilon}_{1}
\cdot\vec{\epsilon}_{2}^{~*})
\alpha_{\pi}+(\vec{k}'\times\vec{\epsilon}_{2}^{~*})
\cdot (\vec{k}\times\vec{\epsilon}_{1})\beta_{\pi}]+\cdots,
\end{equation}
where $T_{B}$ is the Born amplitude
and $\vec{\epsilon}_{1}(\vec{\epsilon}_{2}),\,\omega(\omega')$ and
$\vec{k}(\vec{k}')$ are the polarization vector, energy and momentum
of the initial (final) photon.
Because stable pion targets are unavailable, Compton scattering off
pions has been done indirectly through high-energy pion-nucleus
bremsstrahlung $\pi^{-}Z\rightarrow\pi^{-}Z\gamma$ \cite{Antipov},
radiative pion photoproduction from the proton $\gamma p\rightarrow
\gamma \pi^{+}n$ \cite{Aiber}, and the cross channel two-photon
reaction $\gamma\gamma \rightarrow \pi\pi$ \cite{Boyer, Babusci}.
Recently, a new radiative charged pion photoproduction experiment
has been performed at the Mainz Microtron MAMI \cite{Mainz04}. Their
result differs significantly from the predictions of ChPT:
\begin{equation}(\alpha-\beta)_{\pi^{\pm}}=
(11.6\pm 1.5_{stat}\pm 3.0_{syst}\pm 0.5 _{model})\times 10^{-4}fm^{3}.
\end{equation}
Gasser {\it et al.} \cite{Gasser06} recalculated the two-loop ChPT calculation and obtained
$(\alpha-\beta)_{\pi^{\pm}}=(5.7\pm 1.0)\times 10^{-4}fm^{3}$ with updated values for the LECs at
${\cal O}(p^{4})$ but the result is still in conflict with the MAMI result. Consequently,
the MAMI result fuels the
renewed interest in extracting
the charged pion polarizabilities from the radiative pion photoproduction data.
\newline
\indent
There are mainly two methods to extract the charged pion polarizabilities
$\alpha_{{\pi}^{\pm}}$ and $\beta_{\pi^{\pm}}$ from radiative
charged pion photoproduction.
The method of extrapolation \cite{F85,DF94} is similar to the one
suggested by Chew and Low \cite{ChewLow} in the late 1950's which
has been successfully employed to determine $\pi\pi\rightarrow \pi\pi$ scattering
parameters from the reaction $\pi N\rightarrow \pi\pi N$. However,
this method is based on the assumption that the pion-pole diagram is the
dominant diagram when  $t$, the squared momentum transferred to the
nucleon, is very close to zero.
The authors of \cite{Mainz04} pointed
out that, in the case of radiative charged pion photoproduction,
the pion pole diagram alone is not gauge invariant and one
has to take into account all pion and nucleon pole diagrams.
In order to apply the extrapolation method, one needs not only very
precise data near small $t$ but also good theoretical understanding
of the background (the non pion-pole diagrams). This casts doubt on
the utility of the extrapolation method.
\newline
\indent The second method is to apply some models to calculate the cross section for
the reaction $\gamma p\rightarrow \gamma \pi^{+}n$.
For example, in\cite{Mainz04} two different models were employed.
The first one includes all the pion and nucleon pole diagrams through
the use of the pseudoscalar pion-nucleon coupling.
The second one includes the nucleon and pion pole diagrams (without the
anomalous magnetic moments of the nucleons) and the contributions from the
$\Delta(1232),\,P_{11}(1440),\,D_{13}(1520),\,$ and $S_{11}(1535)$
resonances.
They then determined the value of $(\alpha-\beta)_{\pi^{\pm}}$ by comparing
the predictions of these models to the data.
\newline
\indent
In this article, we explore the possibility of
extracting the charged pion polarizabilities directly from the cross section of the radiative charged pion
photoproduction within the framework of heavy baryon chiral
perturbation theory (HBChPT) (see \cite{BKM} for a review of
HBChPT).
The basic idea here is to calculate the cross section of the reaction
$\gamma p\rightarrow \pi^{+}\gamma n$ by HBChPT then extract
$\alpha_{\pi^{\pm}}$ and $\beta_{\pi^{\pm}}$ from the
experimental data of the cross section.
This approach is essentially model-independent and gauge-invariant.
The complete result of the radiative pion
photoproduction in HBChPT at the one-loop level will be reported
elsewhere\cite{kao08}.
Here we focus on the best experimental setup for the
extraction of the charged pion polarizabilities from the cross section of
radiative charged pion photoproduction.
\newline
\indent
This article is organized as follows. In Sec. II the kinematics of
the radiative pion photoproduction is discussed. In Sec. III we analyze the
amplitude of the radiative pion photoproduction in HBChPT.
The extraction of charged pion polarizabilities from the cross section
of radiative charged pion photoproduction is studied in Sec. IV.
We discuss the applicability
of the extrapolation method in Sec. V.
Several issues are discussed and conclusions are given in Sec.
VI.

\section{Kinematics of radiative charged pion photoproduction}
In this section we discuss the kinematics of radiative charged pion photoproduction.
We adopt the following notations:
\begin{equation}
\gamma(\epsilon_{1},k)+p(P_{1})\rightarrow \gamma
(\epsilon_{2},q)+\pi^{+}(r)+n(P_{2}).
\end{equation}
\indent
Here
$k$=$(\omega_{k},\vec{k}~)$,\,\,$q$=$(\omega_{q},\vec{q}~)$,\,\,
$r$=$(\omega_{r},\vec{r}~)$,
$P_{1}$=$(\sqrt{M_{N}^{2}+|\vec{p}_{1}|^{2}}~,\vec{p}_{1}~),$\,\,\,
$P_{2}$=$(\sqrt{M_{N}^{2}+|\vec{p}_{2}|^{2}}~,\vec{p}_{2}~)$.
$\epsilon_{1}(\epsilon_{2})$ and $k(q)$ are the polarization vector and
momentum of the incoming(outgoing) photon, respectively.
$P_{1}$($P_{2}$) is the momentum of initial(final) nucleon.
We choose the following gauge:
$\epsilon_{1}\cdot v=\epsilon_{2}\cdot v=0,$
where $v$ is the velocity of the nucleon.
The reason to choose this particular gauge is because
the leading order $\gamma NN$ vertex in HBChPT
vanishes in this gauge.
Hence the calculation is significantly simplified.
Furthermore, $\epsilon_{1}\cdot k=\epsilon_{2}\cdot q=0$
because both the incoming and the outgoing photons are real photons.
The most convenient frame is the c.m. frame of the final
$\gamma$-$\pi$ system.
In this frame, one has $\vec{r}+\vec{q}=0$ and the relations as follows:
\begin{eqnarray}S_{1}&\equiv&(r+q)^{2}=
(\omega_{q}+\omega_{r})^{2},\,\,\,\,t\equiv(P_{1}-P_{2})^{2}=
(r-k+q)^{2},\,\,\,\,\cos\theta\equiv\hat{k}\cdot \hat{q},
\nonumber
\\\omega_{q}&=&\frac{S_{1}-m_{\pi}^{2}}{2\sqrt{S_{1}}},\,\,\omega_{k}=
\frac{S_{1}-t}{2\sqrt{S_{1}}},\,\,\,\omega_{r}=
\frac{S_{1}+m_{\pi}^{2}}{2\sqrt{S_{1}}},\,
\omega_{k}-\omega_{q}-\omega_{r}=\frac{-S_{1}-t}{2\sqrt{S_{1}}},
\nonumber
\\u&\equiv&(r-k)^{2}=\frac{1}{2}\left(m_{\pi}^{2}+t-S_{1}+
\frac{tm_{\pi}^{2}}{S_{1}}\right)+\frac{\cos\theta}{2}
\left(m_{\pi}^{2}-S_{1}+t-\frac{tm_{\pi}^{2}}{S_{1}}\right).
\end{eqnarray}
In this article we refer to the plane spanned by $\vec{k}$ and $\vec{r}$ as
the ``scattering plane''.
Note that $S_{1}\ge m_{\pi}^{2}$ and $t\le 0$.

\section{ The Amplitudes of the radiative pion photoproduction in HBChPT }
This section we discuss the amplitudes of radiative charged pion photoproduction in HBChPT.
Before proceeding, one has to determine to which order the
amplitudes need to be computed in HBChPT. Note that HBChPT is
essentially a double expansion, i.e., the combination of a chiral
expansion and a heavy baryon expansion. The amplitudes can be
expressed as:
\begin{equation}{\cal A}=\sum_{n=0,m=0}^{\infty}\frac{{\cal B}_{m,n}}
{(M_{N})^{n}(4\pi F_{\pi})^{2m}}=\sum_{l=2m+n}^{\infty}{\cal A}^{(l)}.
\label{count}
\end{equation}
The predictions for the charged pion polarizabilities Eq.(\ref{ChPT}) are extracted from
the amplitude of Compton scattering of pions at the one loop level, so $\alpha_{\pi^{\pm}}$
and $\beta_{\pi^{\pm}}$ emerge in ${\cal A}^{(2)}(n=0,m=1)$.
The leading-order (LO) amplitude ${\cal A}_{LO}$ is ${\cal
A}^{(0)}(n=0,m=0)$,
and the next-to-leading-order(NLO) amplitude ${\cal A}_{NLO}$ is
${\cal A}^{(1)}(n=1,m=0)$. The next-to-next-to-leading order amplitude
includes ${\cal A}^{(2)}(n=0,m=1)$ and ${\cal A}^{(2)}(n=2,m=0)$.
The square of the amplitude will be expressed as
\begin{eqnarray}
{\cal A}={\cal A}_{LO}+{\cal A}_{NLO}+{\cal
A}_{NNLO}+\cdots\,,\,\,\,\,\,\,\,
|~{\cal A}~|~^{2}&=&|~{\cal A}_{LO}~|~^{2} \nonumber
\\&+&2Re({\cal A}_{LO}{\cal A}_{NLO}^{*}) \nonumber
\\&+&2Re({\cal A}_{LO}{\cal A}_{NNLO}^{*})+|~{\cal A}_{NLO}~|~^{2}+\cdots .
\end{eqnarray}
Hence the cross section can be split into: $\sigma=\sigma_{LO}+\sigma_{NLO}+\sigma_{NNLO}+\cdots$ and
\begin{equation}
\sigma_{LO}\propto |{\cal A}_{LO}|^{2},\,\sigma_{NLO}\propto 2Re({\cal A}_{LO}{\cal A}_{NLO}^{*}) ,\,
\sigma_{NNLO}\propto 2Re({\cal A}_{LO}{\cal A}_{NNLO}^{*})+|~{\cal A}_{NLO}~|~^{2}.
\label{crosssection}
\end{equation}
The charged pion polarizabilities are extracted from $\sigma_{NNLO}$
so that one has to calculate up to the next-to-next-to-leading order in
HBChPT in addition to ${\cal A}^{(0)}(n=0,m=0)$ and
${\cal A}^{(1)}(n=1,m=0)$.

\subsection{Leading and next-to-leading order amplitudes in HBChPT}
\begin{figure}
\epsfysize=2cm\centerline{\epsffile{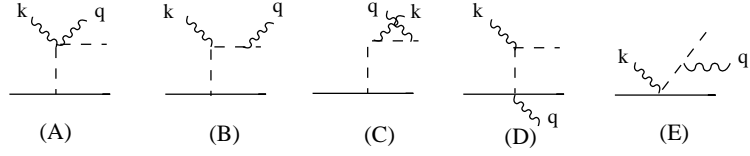}}
\vspace{0.5cm}
\caption{The LO diagrams for radiative pion photoproduction in the $\epsilon_1\cdot v=\epsilon_2\cdot v=0$
gauge. The corresponding amplitude of the diagram (A) is denoted as ${\cal A}_{1A}$ in the text. Similar notations
are applied to the other diagrams. The dotted line represents the pion, the solid line represents the nucleon, the wedged line represents the photon.}
\label{LO}
\end{figure}

The LO diagrams are given in the Fig.(\ref{LO}).
\begin{figure}
\epsfysize=4cm\centerline{\epsffile{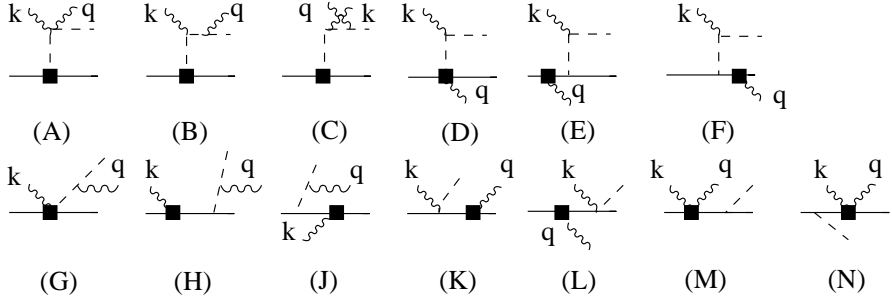}}
\vspace{1cm}
\caption{The NLO diagrams for radiative pion photoproduction in the $\epsilon_1\cdot v=\epsilon_2\cdot v=0$
gauge.
The squares represent the NLO vertices. The corresponding amplitude of the diagram (A) is denoted as ${\cal A}_{2A}$ in the text. Similar notations are applied to the other diagrams. The dotted line represents the pion, the solid line represents the nucleon, the wedged line represents the photon.}
\label{NLO}
\end{figure}
The LO amplitudes for radiative charged pion photoproduction are:
\begin{eqnarray}
{\cal A}_{1A}&=&\frac{-e^{2}g_{A}}{F_{\pi}}[\tau_{c}-\tau_{3}\delta_{c3}]
\frac{(\vec{\epsilon}_{1}\cdot \vec{\epsilon}_{2}^{~*})(\vec{\sigma}\cdot
(\vec{r}-\vec{k}+\vec{q}~))}{t-m_{\pi}^{2}}, \nonumber
\\{\cal A}_{1B}&=&\frac{-2e^{2}g_{A}}{F_{\pi}}[\tau_{c}-\tau_{3}\delta_{c3}]
\frac{(\vec{\epsilon}_{2}^{~*}\cdot\vec{r}~)(\vec{\epsilon}_{1}\cdot
(\vec{r}+\vec{q}~))(\vec{\sigma}\cdot (\vec{r}-\vec{k}+\vec{q}~))}
{[S_{1}-m_{\pi}^{2}][t-m_{\pi}^{2}]}, \nonumber
\\{\cal A}_{1C}&=&\frac{-2e^{2}g_{A}}{F_{\pi}}[\tau_{c}-\tau_{3}\delta_{c3}]
\frac{(\vec{\epsilon}_{1}\cdot \vec{r}~)(\vec{\epsilon}_{2}^{~*}\cdot
(\vec{r}-\vec{k}~))(\vec{\sigma}\cdot
(\vec{r}-\vec{k}+\vec{q}~))}{[u-m_{\pi}^{2}][t-m_{\pi}^{2}]}, \nonumber
\\{\cal A}_{1D}&=&\frac{-e^{2}g_{A}}{F_{\pi}}[\tau_{c}-\tau_{3}\delta_{c3}]
\frac{(\vec{\epsilon}_{1}\cdot \vec{r}~)(\vec{\sigma}\cdot
\vec{\epsilon}_{2}^{~*})}{u-m_{\pi}^{2}}, \nonumber \\{\cal A}_{1E}&=&
\frac{-e^{2}g_{A}}{F_{\pi}}[\tau_{c}-\tau_{3}\delta_{c3}]
\frac{(\vec{\epsilon}_{2}^{~*}\cdot \vec{r}~)(\vec{\sigma}
\cdot \vec{\epsilon}_{1})}{S_{1}-m_{\pi}^{2}}.
\label{eq:LO}
\end{eqnarray}
Here, $c$ is the isospin index of the outgoing pion and * indicates
the complex conjugation.
${\cal A}_{1A}$ represents the amplitude of the diagram (A) in Fig. (\ref{LO}).
Similar notations are applied to other diagrams Fig. (\ref{LO}).
There are several remarks in order regarding
the LO amplitudes. First, they all depend on the nucleon spin.
Second, the diagrams such as (1-B),\,(1-C),\, (1-D), and (1-E)
do not have the corresponding diagrams in $\pi N\rightarrow \pi\pi N$ because there is no
3$\pi$ vertex at leading order. It is important because it
explains the essential difference between $\pi N\rightarrow\pi\pi N$
and $\gamma p\rightarrow \gamma \pi^{+}n$. We will return to this
point in section V when we discuss the applicability of the extrapolation
method. Furthermore, diagrams (1-B)
and (1-E) both vanish in the c.m. frame of the final $\gamma-\pi$
system because $\vec{\epsilon}_{2}^{~*}\cdot\vec{r}=
\vec{\epsilon}_{2}^{~*}\cdot(-\vec{q})=0$. Diagrams (1-C) and (1-D)
vanish if the polarization vector of the incoming photon is
perpendicular to the scattering plane spanned by $\vec{k}$ and
$\vec{r}$. In other words, $\vec{\epsilon}_{1}\cdot \vec{r}=0$.
As a result when the polarization vector of the incoming photon is
perpendicular to the scattering plane the LO amplitude in the c.m. frame of
final $\gamma-\pi$ system becomes ${\cal A}_{1A}$ only. On the other hand,
if the the polarization vector of the incoming photon is
parallel to the scattering plane then the LO amplitude in the c.m. frame of
final $\gamma-\pi$ system becomes ${\cal A}_{1A}+{\cal A}_{1C}+{\cal A}_{1D}$.
\newline
\indent
\begin{figure}
\epsfysize=8cm
\centerline{\epsffile{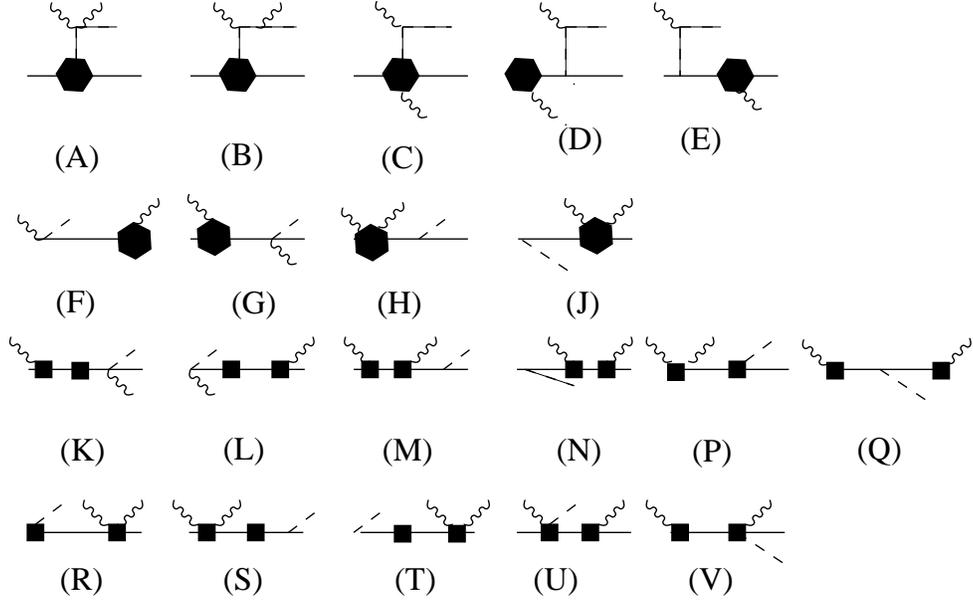}}
\vspace{1cm}
\caption{The NNLO diagrams belonged to ${\cal A}^{(2)}(n=2,m=0)$ for radiative pion electroproduction
in the $\epsilon_1\cdot v=\epsilon_2\cdot v=0$ gauge.
The squares (hexagons) represent the NLO (NNLO) vertices. The dotted line represents the pion, the solid line represents the nucleon, the wedged line represents the photon.}
\label{NNLOtree}
\end{figure}
The NLO diagrams are listed in Fig.(\ref{NLO}).
The NLO amplitudes read:
\begin{equation}
\begin{split}
{\cal
A}_{2A}&=\frac{e^{2}g_{A}}{2M_{N}F_{\pi}}[\tau_{c}-\tau_{3}\delta_{c3}]
\frac{(\vec{\epsilon}_{1}\cdot \vec{\epsilon}_{2}^{~*})(\vec{\sigma}\cdot
(\vec{p}_{1}+\vec{p}_{2}))}{t-m_{\pi}^{2}}\cdot
(\omega_{q}+\omega_{r}-\omega_{k}), \nonumber
\\{\cal
A}_{2B}&=\frac{e^{2}g_{A}}{M_{N}F_{\pi}}[\tau_{c}-\tau_{3}\delta_{c3}]
\frac{(\vec{\epsilon}_{2}^{~*}\cdot \vec{r}~)(\vec{\epsilon}_{1}\cdot
(\vec{r}+\vec{q}~))(\vec{\sigma}\cdot
(\vec{p}_{1}+\vec{p}_{2}))}{[S_{1}-m_{\pi}^{2}][t-m_{\pi}^{2}]}\cdot
(\omega_{q}+\omega_{r}-\omega_{k}), \nonumber
\\{\cal
A}_{2C}&=\frac{e^{2}g_{A}}{M_{N}F_{\pi}}[\tau_{c}-\tau_{3}\delta_{c3}]
\frac{(\vec{\epsilon}_{1}\cdot \vec{r}~)(\vec{\epsilon}_{2}^{~*}\cdot
(\vec{r}-\vec{k}~))(\vec{\sigma}\cdot
(\vec{p}_{1}+\vec{p}_{2}))}{[u-m_{\pi}^{2}][t-m_{\pi}^{2}]}\cdot
(\omega_{q}+\omega_{r}-\omega_{k}), \nonumber
\\{\cal A}_{2D}&=\frac{e^{2}g_{A}}{4M_{N}F_{\pi}}[\tau_{c},\tau_{3}]
\frac{(\vec{\epsilon}_{1}\cdot \vec{r}~)(\vec{\sigma}\cdot
\vec{\epsilon}_{2}^{~*})}{u-m_{\pi}^{2}}\cdot (\omega_{r}-\omega_{k}),
\nonumber
\\{\cal A}_{2E}&=\frac{-e^{2}g_{A}}{2M_{N}F_{\pi}}
[\tau_{c}\tau_{3}-\delta_{c3}][1+\tau_{3}]\frac{(\vec{\epsilon}_{1}\cdot
\vec{r}~)(\vec{\sigma}\cdot (\vec{r}-\vec{k}~))(\vec{\epsilon}_{2}^{~*}\cdot
\vec{p}_{1})}{u-m_{\pi}^{2}}\cdot \frac{1}{\omega_{q}}, \nonumber
\\&-\frac{e^{2}g_{A}}{8M_{N}F_{\pi}}[\tau_{c}\tau_{3}-\delta_{c3}]\tilde{\mu}
\frac{(\vec{\epsilon}_{1}\cdot \vec{r}~)(\vec{\sigma}\cdot
(\vec{r}-\vec{k}~))
[\vec{\sigma}\cdot \vec{\epsilon}_{2}^{~*},\vec{\sigma}\cdot
\vec{q}~]}{u-m_{\pi}^{2}}\cdot \frac{1}{\omega_{q}}, \nonumber
\\{\cal A}_{2F}&=\frac{e^{2}g_{A}}{2M_{N}F_{\pi}}[1+\tau_{3}]
[\tau_{c}\tau_{3}-\delta_{c3}]\frac{(\vec{\epsilon}_{1}\cdot
\vec{r})(\vec{\sigma}\cdot (\vec{r}-\vec{k}~))(\vec{\epsilon}_{2}^{~*}\cdot
\vec{p}_{2})}{u-m_{\pi}^{2}}\cdot \frac{1}{\omega_{q}}, \nonumber
\\&+\frac{e^{2}g_{A}}{8M_{N}F_{\pi}}\tilde{\mu}[\tau_{c}\tau_{3}-\delta_{c3}]
\frac{(\vec{\epsilon}_{1}\cdot \vec{r}~)[\vec{\sigma}\cdot
\vec{\epsilon}_{2}^{~*},\vec{\sigma}\cdot \vec{q}~](\vec{\sigma}\cdot
(\vec{r}-\vec{k}~))}{u-m_{\pi}^{2}}\cdot \frac{1}{\omega_{q}}, \nonumber
\\{\cal A}_{2G}&=\frac{e^{2}g_{A}}{4M_{N}F_{\pi}}[\tau_{c},\tau_{3}]
\frac{(\vec{\epsilon}_{2}^{~*}\cdot \vec{r}~)(\vec{\sigma}\cdot
\vec{\epsilon}_{1})}{S_{1}-m_{\pi}^{2}}\cdot (\omega_{r}+\omega_{q}),
\nonumber
\\{\cal
A}_{2H}&=\frac{e^{2}g_{A}}{2M_{N}F_{\pi}}[\tau_{c}\tau_{3}-\delta_{c3}]
[1+\tau_{3}]\frac{(\vec{\epsilon}_{2}^{~*}\cdot \vec{r}~)(\vec{\sigma}\cdot
(\vec{r}+\vec{q}))(\vec{\epsilon}_{1}\cdot\vec{p}_{1})}{S_{1}-m_{\pi}^{2}}\cdot
\frac{1}{\omega_{k}}, \nonumber \\&+\frac{e^{2}g_{A}}{8M_{N}F_{\pi}}
[\tau_{c}\tau_{3}-\delta_{c3}]\tilde{\mu}\frac{(\vec{\epsilon}_{2}^{~*}\cdot
\vec{r}~)(\vec{\sigma}\cdot (\vec{r}+\vec{q}~))[\vec{\sigma}\cdot
\vec{\epsilon}_{1},\vec{\sigma}\cdot \vec{k}~]}{S_{1}-m_{\pi}^{2}}\cdot
\frac{1}{\omega_{k}}, \nonumber \\{\cal A}_{2J}&=
\frac{-e^{2}g_{A}}{2M_{N}F_{\pi}}[1+\tau_{3}][\tau_{c}\tau_{3}-\delta_{c3}]
\frac{(\vec{\epsilon}_{2}^{~*}\cdot \vec{r}~)(\vec{\sigma}\cdot
(\vec{r}+\vec{q}~))(\vec{\epsilon}_{1}\cdot \vec{p}_{2})}{S_{1}-m_{\pi}^{2}}
\cdot \frac{1}{\omega_{k}}, \nonumber \\&-\frac{e^{2}g_{A}}{8M_{N}F_{\pi}}
\tilde{\mu}[\tau_{c}\tau_{3}-\delta_{c3}]\frac{(\vec{\epsilon}_{2}^{~*}\cdot
\vec{r})[\vec{\sigma}\cdot \vec{\epsilon}_{1},\vec{\sigma}\cdot \vec{k}~]
(\vec{\sigma}\cdot (\vec{r}+\vec{q}~))}{S_{1}-m_{\pi}^{2}}\cdot
\frac{1}{\omega_{q}}, \nonumber
\\{\cal A}_{2K}&=\frac{-e^{2}g_{A}}{4M_{N}F_{\pi}}[1+\tau_{3}]
[\tau_{c}\tau_{3}-\delta_{c3}]\frac{(\vec{\epsilon}_{2}^{~*}\cdot
\vec{p}_{2})(\vec{\sigma}\cdot \vec{\epsilon}_{1})}{\omega_{q}}
\nonumber
\\&-\frac{e^{2}g_{A}}{16M_{N}F_{\pi}}\tilde{\mu}[\tau_{c}\tau_{3}-\delta_{c3}]
[1+\tau_{3}]\frac{[\vec{\sigma}\cdot\vec{\epsilon}_{2}^{~*},\vec{\sigma}\cdot
\vec{q}~](\vec{\sigma}\cdot \vec{\epsilon}_{1})}{\omega_{q}},
\nonumber
\\{\cal A}_{2L}&=\frac{e^{2}g_{A}}{4M_{N}F_{\pi}}
[\tau_{c}\tau_{3}-\delta_{c3}][1+\tau_{3}]\frac{(\vec{\epsilon}_{1}\cdot
\vec{p}_{1})(\vec{\sigma}\cdot
\vec{\epsilon}_{2}^{~*})}{\omega_{q}+\omega_{r}}
\nonumber
\\&+\frac{e^{2}g_{A}}{16M_{N}F_{\pi}}[\tau_{c}\tau_{3}-\delta_{c3}]
\tilde{\mu}\frac{[\vec{\sigma}\cdot\vec{\epsilon}_{1},\vec{\sigma}\cdot
\vec{k}~](\vec{\sigma}\cdot
\vec{\epsilon}_{2}^{~*})}{\omega_{q}+\omega_{r}},
\nonumber
\\{\cal A}_{2M}&=\frac{-e^2 g_{A}}{4M_{N}F_{\pi}}\tau_{c}(1+\tau_{3})
\frac{(\vec{\sigma}\cdot\vec{r}~)(\vec{\epsilon}_{1}\cdot
\vec{\epsilon}_{2}^{~*})}{\omega_{r}},\nonumber \\{\cal A}_{2N}&=
\frac{e^2 g_{A}}{4M_{N}F_{\pi}}(1+\tau_{3})\tau_{c}\frac{(\vec{\sigma}
\cdot\vec{r}~)(\vec{\epsilon}_{1}\cdot\vec{\epsilon}_{2}^{~*})}{\omega_{r}},
\label{eq:NLO}
\end{split}
\end{equation}
where $\tilde{\mu}=(1+\kappa_{s})+\tau_{3}(1+\kappa_{v})$ and
$\kappa_{v}(\kappa_{s})=3.76(-0.120)$ is the isovector (isoscalar) anomalous
magnetic moment of the nucleon.
${\cal A}_{2A}$ represents the amplitude of the diagram (A) in Fig. (\ref{NLO}).
Similar notations are applied to other diagrams Fig. (\ref{NLO}).
In the c.m frame of the final $\gamma$-$\pi$ system,
the diagrams (2-B),(2-G), (2-H), and (2-J) also vanish because $\vec{\sigma}\cdot (\vec{r}+\vec{q})=0$ and
$\vec{\epsilon}_{2}^{~*}\cdot\vec{r}=\vec{\epsilon}_{2}^{~*}\cdot(-\vec{q})=0$.
Furthermore,
the diagrams (2-C), (2-D), (2-E), and (2-F) vanish if the polarization vector of
the incoming photon is perpendicular to the scattering plane because $\vec{\epsilon}_{1}\cdot \vec{r}=0$.
As a result when the polarization vector of the incoming photon is
perpendicular to the scattering plane the NLO amplitude in the c.m. frame of
final $\gamma-\pi$ system becomes
\begin{equation}
{\cal A}_{NLO}^{\perp}={\cal A}_{2A}+{\cal A}_{2K}+{\cal A}_{2L}+{\cal A}_{2M}+{\cal A}_{2N},
\end{equation}
which contains no term proportional to $1/(u-m_{\pi}^{2})$. As a matter of fact, in this particular case
the sum of LO and NLO amplitudes is free of the pole at $u=m_{\pi}^{2}$.
This is our first important observation.

\subsection{Next-to-next leading order amplitudes in HBChPT}
\begin{figure}
\epsfysize=14.2cm
\centerline{\epsffile{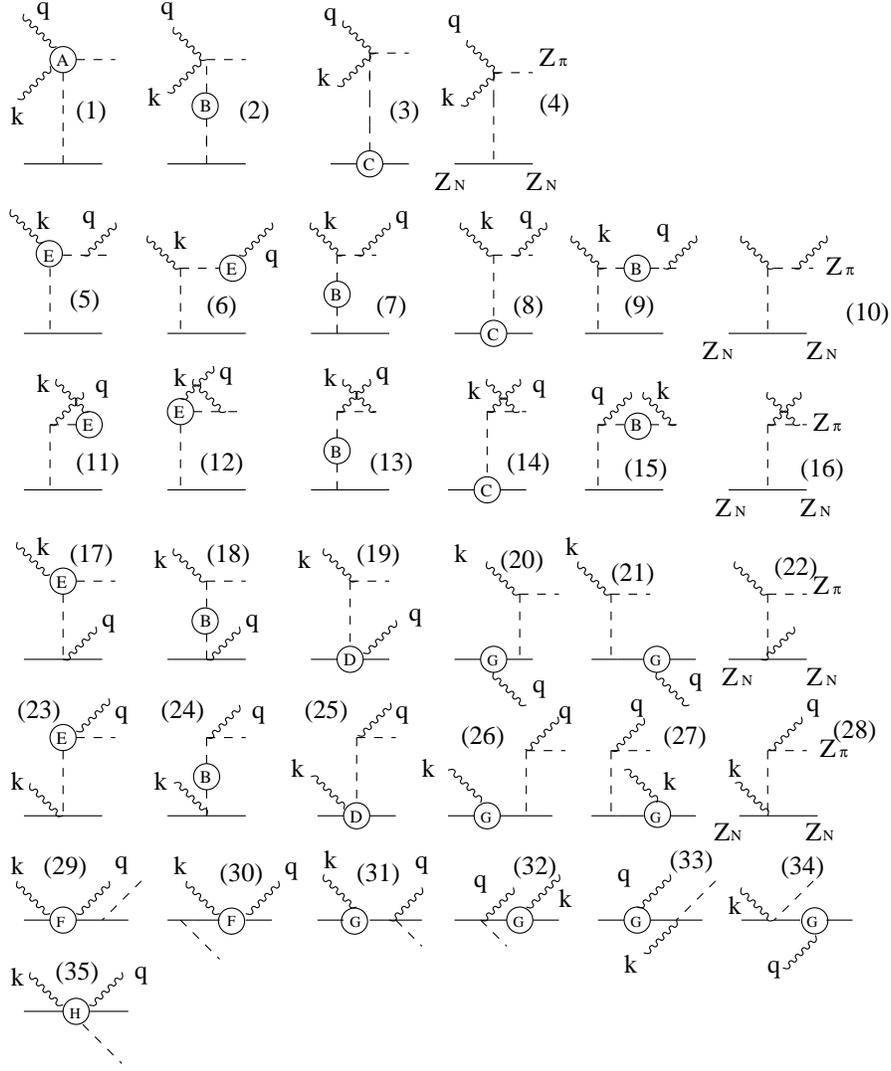}}
\vspace{1cm}
\caption{The NNLO diagrams belonged to ${\cal A}^{(2)}(n=0,m=1)$ for radiative pion electroproduction in the
$\epsilon_1\cdot v=\epsilon_2\cdot v=0$
gauge. The bubbles labelled ${\cal M}_{A}$ -- ${\cal M}_{H}$ represent the one-loop chiral corrections to the sub-processes. The dotted line represents the pion, the solid line represents the nucleon, the wedged line represents the photon.}
\label{sut}
\end{figure}

The NNLO amplitude includes ${\cal A}^{(2)}(n=2,m=0)$ and
${\cal A}^{(2)}(n=0,m=1)$.
The diagrams which contribute to ${\cal A}^{(2)}(n=2,m=0)$ are given by
Fig.(\ref{NNLOtree});
the diagrams which contribute to  ${\cal A}^{(2)}(n=0,m=1)$ are given by
Fig(\ref{sut}).

\begin{figure}
\epsfysize=12cm
\centerline{\epsffile{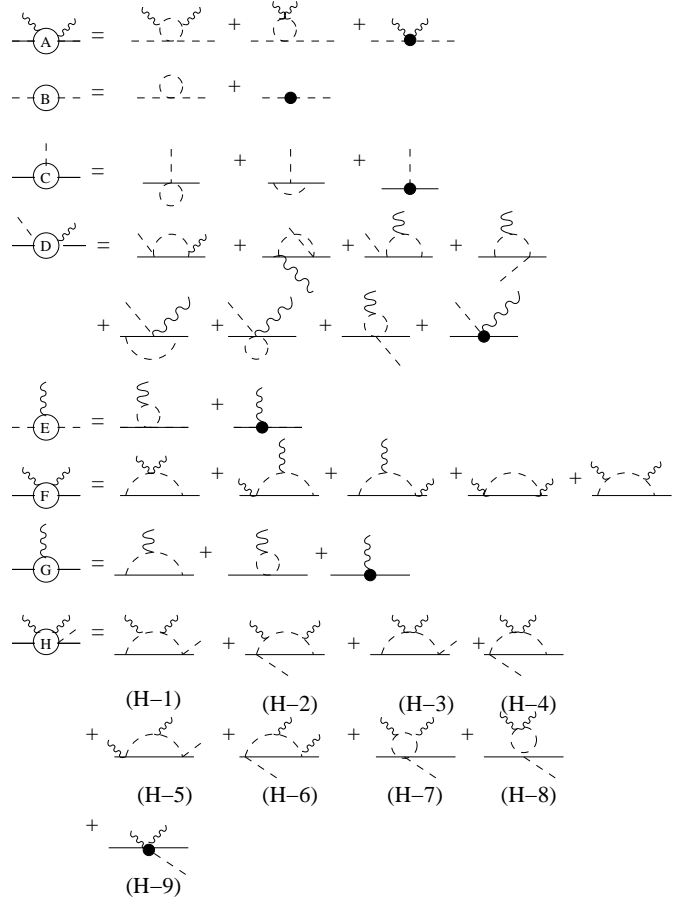}}
\vspace{1cm}
\caption{The diagrams denoted as ${\cal M}_{A}$ --${\cal M}_{H}$ in the text
represent the one-loop chiral corrections to the sub-processes.
The detail explanation of each diagram is given in the text.
The full circle represents the vertex from ${\cal L}_{\pi N}^{(3)}$ or
${\cal L}_{\pi\pi}^{(4)}$. The dotted line represents the pion, the solid line represents the nucleon, the wedged line represents the photon.}
\label{bubble}
\end{figure}
The ``bubbles'' appearing in the diagrams of Fig (\ref{sut}), denoted as
from ${\cal M}_{A}$ to ${\cal M}_{H}$, are the sums of the
one-particle irreducible diagrams of
some sub-processes.
An explicit graphic explanation
of each ``bubbles'' is given at Fig(\ref{bubble}). They are the
one-loop chiral corrections to the
tree-level amplitudes of the sub-processes with at least one off-shell legs.
(Except ${\cal M}_{H}$ in which the five legs are all on-shell).
They can be calculated in HBChPT and, actually, most of them have been
calculated before. However these calculations have done under different
different definition of the pion and nucleon fields. The complete calculation
of all sub-diagrams under the same definition of the fields is left for
future publication \cite{kao08}.
$Z_{\pi}$ and $Z_{N}$ are the wave function renormalization factors for the
pion and nucleon, respectively. Both of them can be found in the literatures \cite{Fettes00}
${\cal M}_{A}(\gamma+\pi^{+}\rightarrow \gamma+\pi^{+})$ represents one-loop chiral
contribution to Compton scattering from the virtual incoming pion.
If the incoming photon is also virtual, then this sub-diagram also carry
the information of the so-called generalized polarizabilities as studied in
\cite{UOFMS02, GP}.
${\cal M}_{B}(\pi\rightarrow \pi)$ represents the one-loop chiral correction
to the pion mass.
${\cal M}_{C}(p\rightarrow \pi^{+}+n)$ represents the one-loop chiral correction
to the axial form factor of the nucleon.
${\cal M}_{D}(\pi^{-}+p\rightarrow \gamma+n)$ is the one-loop chiral correction
to the amplitude of the radiative capture of the virtual
charged pion\cite{Pioncapture}.
${\cal M}_{E}(\pi^{+}\rightarrow \gamma+\pi^{+})$ is the one-loop chiral correction to
the pion electromagnetic form factor.
${\cal M}_{F}(\gamma+N\rightarrow \gamma+N)$ represents Compton
scattering from an off-shell nucleon (either incoming or outgoing)
\cite{Compton} at one-loop level
and ${\cal M}_{G}(N\rightarrow \gamma+N)$ stands for the one-loop chiral correction to the
nucleon electromagnetic form factor.
The only sub-diagram that has never been calculated in HBChPT is
${\cal M}_{H}(\gamma+p\rightarrow\pi^{+}+\gamma+n)$.
The amplitudes contributing to ${\cal M}_{H}$ are quite lengthy and are
given in the appendix.
\begin{figure}
\epsfysize=3cm
\centerline{\epsffile{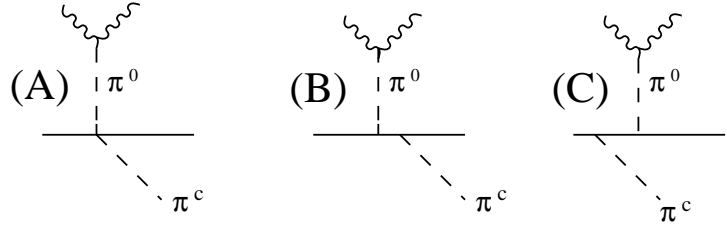}}
\vspace{1cm}
\caption{The diagrams for the radiative pion electroproduction in the
temporal gauge with the WZW vertex. The index c is isospin index.}
\label{diagram:wzw}
\end{figure}
Besides, there is one exceptional class of diagrams which contain the
Wess-Zumino-Witten anomalous term $\pi^{0}\rightarrow 2\gamma$
\cite{WZW}(see Fig.(\ref{diagram:wzw})).
The WZW term is the consequence of the chiral anomaly of QCD \cite{ABJ}.
These diagrams contribute to the NNLO amplitudes:
\begin{eqnarray}
{\cal A}^{WZW}_{A}&=&i\frac{e^{2}}{2(\pi F_{\pi})^{2}F_{\pi}}
[\tau_{3},\tau_{c}]\epsilon_{\mu\nu\alpha\beta}k^{\mu}
\epsilon_{1}^{\nu}q^{\alpha}\epsilon_{2}^{*\beta}
\frac{1}{(q-k)^{2}-m_{\pi}^{2}}\cdot
(\omega_{k}-\omega_{q}+\omega_{r}), \nonumber
\\{\cal A}^{WZW}_{B}&=&-\frac{e^{2}g_{A}^{2}}
{(4\pi F_{\pi})^{2}F_{\pi}}\tau_{3}\tau_{c}
\epsilon_{\mu\nu\alpha\beta}k^{\mu}\epsilon_{1}^{\nu}q^{\alpha}
\epsilon_{2}^{*\beta}\frac{(\vec{\sigma}\cdot (\vec{q}-\vec{k}))
(\vec{\sigma}\cdot \vec{r})}{(q-k)^{2}-m_{\pi}^{2}}\frac{1}{\omega_{r}},
\nonumber
\\{\cal A}^{WZW}_{C}&=&\frac{e^{2}g_{A}^{2}}{(4\pi F_{\pi})^{2}F_{\pi}}
\tau_{c}\tau_{3}\epsilon_{\mu\nu\alpha\beta}k^{\mu}
\epsilon_{1}^{\nu}q^{\alpha}\epsilon_{2}^{*\beta}\frac{(\vec{\sigma}\cdot
\vec{r})(\vec{\sigma}\cdot (\vec{q}-\vec{k}))}{(q-k)^{2}-m_{\pi}^{2}}
\frac{1}{\omega_{r}}, \nonumber \\{\cal A}^{WZW}_{B}+{\cal A}^{WZW}_{C}&=&
\frac{e^{2}g_{A}^{2}}{(4\pi F_{\pi})^{2}F_{\pi}}
\epsilon_{\mu\nu\alpha\beta}k^{\mu}\epsilon_{1}^{\nu}q^{\alpha}
\epsilon_{2}^{*\beta}\nonumber \\&&\cdot
\frac{1}{\omega_{r}}\left\{[\tau_{c},\tau_{3}]\frac{\vec{r}\cdot
(\vec{q}-\vec{k})}{(q-k)^{2}-m_{\pi}^{2}}-2\delta_{c3}
\frac{i\vec{\sigma}\cdot (\vec{q}-\vec{k})\times\vec{r}}
{(q-k)^{2}-m_{\pi}^{2}}\right\}.
\label{wzw}
\end{eqnarray}
Note that $(q-k)^2-m_{\pi}^{2}=t-u-S_{1}$.
By conservation of energy, one obtains the relation for small momentum
transfer,
\begin{equation}\omega_{k}-\omega_{q}-\omega_{r}=
\sqrt{M_{N}^{2}+|\vec{p}_{2}|^{2}}-\sqrt{M_{N}^{2}+
|\vec{p}_{1}|^{2}}=\frac{|\vec{p}_{1}|^2-|\vec{p}_{2}|^2}
{2M_{N}}+{\cal O}\left(\frac{1}{M_{N}^{2}}\right).
\end{equation}
Hence ${\cal A}^{WZW}_{A}$ is suppressed.
On the other hand, ${\cal A}^{WZW}_{B}+{\cal A}^{WZW}_{C}$ is spin-independent for
the charged pion.
Since ${\cal A}_{LO}$ is spin-dependent amplitude,
only the spin-dependent amplitude of ${\cal A}_{NNLO}$
will contribute
to the cross section in Eq.~(\ref{crosssection}) because one has to sum over
the initial nucleon spin if
the proton target is unpolarized. The product of one spin-dependent amplitude
and another spin-independent amplitude is spin-dependent and it will vanish after summing over
the spin.
As one can see from Eq.~(\ref{wzw}) the total amplitude of the leading order
WZW-type diagrams is spin-independent so that the WZW type diagrams
do not contribute to the cross section in
Eq.~(\ref{crosssection}).

\section{Extracting charged pion polarizabilities from the cross section of radiative charged pion photoproduction}
\subsection{Chiral Lagrangian and the counter terms}
In this section we discuss how to extract $\alpha_{{\pi}^{\pm}}$ and $\beta_{{\pi}^{\pm}}$
from the cross section for radiative charged pion photoproduction in HBChPT.
According to Eq.~(\ref{crosssection}), the cross section depends on the
charged pion
polarizabilities through the interference term between ${\cal A}_{LO}$ and
the amplitude of the diagram (1) in Fig. (\ref{sut}).
Since there are no unknown parameters in the LO and NLO amplitudes,
the main uncertainty in this approach comes from the other unknown parameters
in the NNLO amplitude.
They are
the so-called low energy constants
(LECs), which are the coefficients
of the counter terms in the chiral Lagrangian. In principle their values
can be determined only through the experimental data.
Note that by studying the fixed point structure of renormalization group equations,
the ratios of some LECs can be estimated \cite{Kim}.
The chiral Lagrangian is expanded as:
\begin{equation}{\cal L}_{eff}={\cal L}_{\pi\pi}^{(2)}+
{\cal L}_{\pi\pi}^{(4)}+{\cal L}_{\pi N}^{(1)}+
{\cal L}_{\pi N}^{(2)}+{\cal L}_{\pi N}^{(3)}+\cdots.
\end{equation}
The charged pion polarizabilities $\alpha_{\pi^{\pm}}$ and  $\beta_{\pi^{\pm}}$
are the LECs in ${\cal L}_{\pi\pi}^{(4)}$.
There are seven LECs in ${\cal L}_{\pi N}^{(2)}$ but only two are involved
in our calculation and they are just the isovector and isoscalar
anomalous magnetic moments of the nucleon, $\kappa_{s}$ and $\kappa_{v}$.
${\cal M}_{B}$ contains two LECs in ${\cal L}_{\pi\pi}^{(4)}$ but they
are absorbed into the renormalized pion mass.
Similarly, ${\cal M}_{C}$ also contains two LECs but they are both
absorbed into the axial coupling constant $g_{A}$ and the pion decay
constant $F_{\pi}$.
Consequently, we only need consider two sub-diagrams, ${\cal M}_{H}$
and ${\cal M}_{D}$.
${\cal M}_{H}$ contains a $\gamma\gamma \pi NN$ vertex in the diagram (H-9)
in Fig. (\ref{bubble}).
This vertex is from the following terms of ${\cal L}_{\pi N}^{(3)}$ \cite{p3L}:
\begin{eqnarray}
{\cal L}_{\pi
N}^{(3)}&=&\bar{N}\left[d_{8}\epsilon^{\mu\nu\alpha\beta}
\langle\tilde{F}_{\mu\nu}^{+}u_{\alpha} \rangle
v_{\beta}+d_{9}\epsilon^{\mu\nu\alpha\beta}\langle
\tilde{F}_{\mu\nu}^{+}\rangle u_{\alpha}v_{\beta} \right. \nonumber \\
&+&\left.
d_{20}iS^{\mu}v^{\nu}[\tilde{F}_{\mu\nu}^{+},v\cdot u]+
d_{21}iS^{\mu}[\tilde{F}_{\mu\nu}^{+},u^{\nu}]+d_{22}S^{\mu}
[D^{\nu},\tilde{F}_{\mu\nu}^{-}]\right]N+\cdots
\label{L}
\end{eqnarray}
where $u^{2}=U=\sqrt{1-\pi^{2}/F_{\pi}^{2}}+i\vec{\tau}\cdot\vec{\pi},
\,\,\,u_{\mu}=i(u^{\dagger}D_{\mu}u-uD_{\mu}u^{\dagger})$,
$\langle A\rangle\equiv Tr(A)$,
$F_{\mu\nu}^{\pm}=e(\partial_{\mu}A_{\nu}-\partial_{\nu}A_{\mu})
(uQu^{\dagger}\pm u^{\dagger}Qu)$, $\tilde{F}_{\mu\nu}^{\pm}=
F_{\mu\nu}^{\pm}-\frac{1}{2}\langle F_{\mu\nu}^{\pm}\rangle$,
$S_{\mu}=\frac{i}{2}\gamma_{5}\sigma_{\mu\nu}v^{\nu}=
(0,\frac{1}{2}\vec{\sigma}~)$
where $v_{\mu}$ is taken as $(1,\vec{0}~)$.
The values of the LECs $d_{i}$ are determined by experimental data.
The terms with $d_{8}$ and $d_{9}$ are independent of the nucleon spin.
They will not contribute to the cross section as long as the proton target is unpolarized.
The term with $d_{20}$ vanishes in the gauge $\epsilon_1\cdot v=\epsilon_2\cdot v=0$.
Therefore, the amplitude of (H-9) is given as
\begin{equation}
{\cal A}^{c.t.}_{H}=\frac{e^{2}}{F_{\pi}}\left(d_{21}+
\frac{d_{22}}{2}\right)[\tau_{c}-\delta_{c3}\tau_{3}]
[(\vec{\epsilon}_{1}\cdot\vec{\epsilon}_{2}^{~*})
(\vec{\sigma}\cdot (\vec{k}-\vec{q})~)-(\vec{\sigma}\cdot \vec{\epsilon_{1}})
(\vec{\epsilon}_{2}^{~*}\cdot \vec{k}~)+
(\vec{\sigma}\cdot\vec{\epsilon}_{2}^{~*})(\vec{\epsilon}_{1}\cdot
\vec{q}~)].
\label{cta}
\end{equation}
Another sub-diagram containing unknown parameters is ${\cal M}_{D}$ which
is the amplitude for  $\pi^{+}+p\rightarrow \gamma+n$,
where all external legs are on-shell except the $\pi^{+}$.
This sub-diagram includes one $\gamma\pi NN$ vertex and the corresponding
amplitude is
\begin{eqnarray}
&&{\cal A}_{D}^{c.t}=\frac{-2e^{2}g_{A}}{F_{\pi}}
[\tau_{c}-\tau_{3}\delta_{c3}]\left\{\frac{\vec{\epsilon}_{1}\cdot
\vec{r}}{u-m_{\pi}^{2}}\left[d_{20}\cdot\omega_q^{2}(\vec{\sigma}\cdot
\vec{\epsilon}_{2}^{~*}) \right.\right.\nonumber \\
&+&\left.\left. (d_{21}+\frac{d_{22}}{2})[(\vec{\sigma}\cdot
\vec{\epsilon}_{2}^{~*})(\omega_q^{2}+\vec{q}\cdot
(\vec{r}-\vec{k}~))-(\vec{\sigma}\cdot \vec{q}~)
(\vec{\epsilon}_{2}^{~*}\cdot (\vec{r}-\vec{k}~))]~\right]\right. \nonumber
\\&+&\left. \frac{\vec{\epsilon}_{2}^{~*}\cdot \vec{r}}{S_{1}-m_{\pi}^{2}}
\left[d_{20}\cdot\omega_{k}^{2}(\vec{\sigma}\cdot \vec{\epsilon_{1}}) +(d_{21}+
\frac{d_{22}}{2})[(\vec{\sigma}\cdot
\vec{\epsilon_{1}})(\omega_{k}^{2}-\vec{k}
\cdot(\vec{r}+\vec{q}~))-(\vec{\sigma}\cdot \vec{k}~)(\vec{\epsilon}_{1}
\cdot (\vec{r}+\vec{q}~))]~\right]\right\}.\nonumber \\
\label{ctD}
\end{eqnarray}
Note that the combination $d_{21}+\frac{d_{22}}{2}$ appears in the
$\pi\gamma NN$ vertex is the same as the one in the
$\pi\gamma\gamma NN$ vertex
because the latter is simply the minimal substitution of the former one.
The values of $d_{20}$ and $d_{21}+\frac{d_{22}}{2}$ are determined by the experimental data of charged pion
photoproduction and/or radiative pion capture.
They also play important roles in the nucleon spin polarizability
$\gamma_{0}$
at the two-loop level \cite{kaoM}.
Their contributions to the cross section of radiative charged pion photoproduction
are the main theoretical uncertainties of the approach within HBChPT framework.

\subsection{The numerical results}
\begin{figure}
\epsfysize=8cm
\centerline{\epsffile{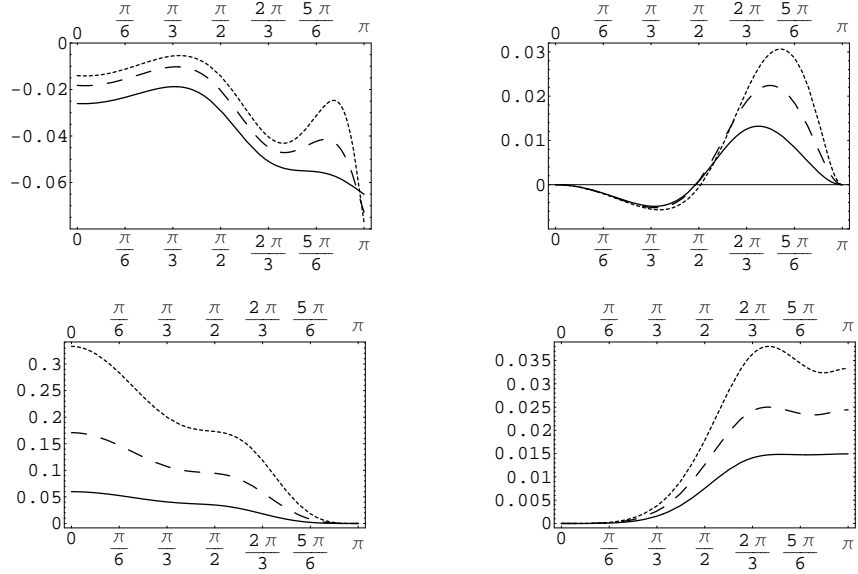}}
\vspace{1cm}
\caption{$R$'s as functions of $\theta$. $R_{\xi}$(left) and
$R_{\eta}$(right)
are in the upper panel.$R_{+}$(left) and $R_{-}$(right) are in the lower
panel.
The solid (dashed, dotted) line is case of $S_{1}$=$4 m_{\pi}^{2}$
($7 m_{\pi}^{2}$, $10 m_{\pi}^{2}$).
$t$=$-3 m_{\pi}^{2}$ in all cases.}
\label{U}
\end{figure}

The previous section shows that total cross section for radiative photoproduction depends not
only on charged pion polarizabilities $\alpha_{\pi^{\pm}}$ and $\beta_{\pi^{\pm}}$,
but also on the LECs $d_{20}$ and $d_{21}+\frac{d_{22}}{2}$:
\begin{eqnarray}
\frac{d^{3}\sigma_{\gamma N\rightarrow\gamma\pi N}}{dtdS_{1}d{\Omega}}&=&
\left(\frac{d^{3}\sigma}{dtdS_{1}d{\Omega}}\right)^{LO}+
\left(\frac{d^{3}\sigma}{dtdS_{1}d{\Omega}}\right)^{NLO}+
\left(\frac{d^{3}\sigma}{dtdS_{1}d{\Omega}}\right)^{NNLO}_{0}\nonumber
\\&+&(\tilde{\alpha}+\tilde{\beta})\cdot\left(\frac{d^{3}\sigma}
{dtdS_{1}d{\Omega}}\right)^{NNLO}_{+}+(\tilde{\alpha}-\tilde{\beta})
\cdot\left(\frac{d^{3}\sigma}{dtdS_{1}d{\Omega}}\right)^{NNLO}_{-}\nonumber
\\&+&\eta\cdot\left(\frac{d^{3}\sigma}{dtdS_{1}d{\Omega}}\right)^{NNLO}_{\eta}
+\xi\cdot\left(\frac{d^{3}\sigma}{dtdS_{1}d{\Omega}}\right)^{NNLO}_{\xi}.
\label{cros}
\end{eqnarray}
Here we have defined dimensionless quantities whose magnitudes are
${\cal O}(1)$:
$$\tilde{\alpha}\pm\tilde{\beta}\equiv(\alpha_{\pi^{\pm}}\pm\beta_{\pi^{\pm}})
\cdot \frac{(4\pi F_{\pi})^{2}m_{\pi}}{\alpha_{em}},\,\,\,\,\xi\equiv
\left(d_{21}+\frac{d_{22}}{2}\right)\cdot(4\pi F_{\pi})^{2},\,\,\,\eta\equiv
\left(d_{20}+d_{21}+\frac{d_{22}}{2}\right)\cdot(4\pi F_{\pi})^{2}.$$
To extract the charged pion polarizabilities, one should look for the experimental
configuration in which both of $\xi$ and $\eta$ have the least impacts on
the cross section.
At the same time, one should also seek the configuration which makes the
cross section most sensitive to the values of $\alpha_{\pi^{\pm}}$ and
$\beta_{\pi^{\pm}}$.
Hence we define the following dimensionless quantities:
\begin{eqnarray}R_{+}&=&\left(\frac{d^{3}\sigma}{dtdS_{1}d{\Omega}}
\right)^{NNLO}_{+}/\left(\frac{d^{3}\sigma}{dtdS_{1}d{\Omega}}\right)^{LO},
\,\,R_{-}=\left(\frac{d^{3}\sigma}{dtdS_{1}d{\Omega}}\right)^{NNLO}_{-}/
\left(\frac{d^{3}\sigma}{dtdS_{1}d{\Omega}}\right)^{LO},\,\,\nonumber
\\R_{\eta}&=&\left(\frac{d^{3}\sigma}{dtdS_{1}d{\Omega}}\right)^{NNLO}_{\eta}
/\left(\frac{d^{3}\sigma}{dtdS_{1}d{\Omega}}\right)^{LO},\,\,R_{\xi}=
\left(\frac{d^{3}\sigma}{dtdS_{1}d{\Omega}}\right)^{NNLO}_{\xi}/
\left(\frac{d^{3}\sigma}{dtdS_{1}d{\Omega}}\right)^{LO}.
\end{eqnarray}
The configuration with the smaller values of $R_{\xi}$
and $R_{\eta}$ is preferred because it means the effects of $\xi$ and
$\eta$ are smaller. At the same time, the experimental setup which gives the larger values
of $R_{+},\,\,R_{-}$ is also preferred since in this case the effects of
the charged pion polarizabilities in the cross section are more pronounced.
Therefore one should look for the experimental setup with small $R_{\xi}$ and
$R_{\eta}$ and large $R_{+}$ or $R_{-}$.
\newline\indent
When the incoming photon is unpolarized,
from Fig.(\ref{U})
we observe that $R_{+}$ is large at forward angles and $R_{-}$ is large at
backward angles. Moreover, both $R_{+}$ and $R_{-}$ increase when $S_{1}$
increases. Although $R_{+}$ at forward angles is about 10 times larger than $R_{-}$
at backward angles,
$\alpha_{\pi^{\pm}}+\beta_{\pi^{\pm}}$ is expected to be far smaller than
$\alpha_{\pi^{\pm}}-\beta_{\pi^{\pm}}$.
(According to Eq.(\ref{p6}) $\alpha_{\pi^{\pm}}-\beta_{\pi^{\pm}}$ is about 10
times larger than $\alpha_{\pi^{\pm}}+\beta_{\pi^{\pm}}$).
Therefore, their effects are expected to be of the same magnitude.
\newline
\indent
Now we turn to the values of $R_{\xi}$ and $R_{\eta}$. It is interesting to see
the behaviours of $R_{\xi}$ and $R_{\eta}$ are very different.
$R_{\xi}$ is small and insensitive to $\theta$ at forward angles,
but it becomes very sensitive to $\theta$ at backward angles. Its absolute
value increases dramatically in the region $90^{\circ}\le\theta\le 120^{\circ}$
then drops in the range $120^{\circ}\le\theta\le 150^{\circ}$, and increases again
in the range of $150^{\circ}\le\theta \le 180^{\circ}$.
$R_{\eta}$ is very small at forward angles, increases rapidly between
$90^{\circ}\le\theta\le 150^{\circ}$, then drops at very backward angles.
\newline
\indent
In order to extract $\alpha_{\pi^{\pm}}-\beta_{\pi^{\pm}}$, one has to
measure the cross section in the range $180^{\circ}\ge \theta\ge 120^{\circ}$ with large $S_{1}$
because $R_{-}$ is large under such conditions.
However, the effect of $\xi$ is most pronounced at very backward angles.
In particular, the $\theta$ dependence of $R_{\xi}$ is complicated
at large $S_{1}$ and extreme backward angles.
Therefore, one should avoid the region $\theta\ge 150^{\circ}$.
But, even in the region $120^{\circ}\le\theta\le 150^{\circ}$,  $R_{\xi}$ and
$R_{\eta}$ are both comparable to $R_{-}$.
We conclude that it is necessary to take the effects of $\xi$ and $\eta$
into consideration when one extract $\alpha_{\pi^{\pm}}-\beta_{\pi^{\pm}}$
from the cross section of radiative charged pion photoproduction.
At forward angles,
the effect of $\eta$ is quite small but the effect of $\xi$ is still
comparable to the effect of $\alpha_{\pi^{\pm}}+\beta_{\pi^{\pm}}$.
Therefore, one should fit $\eta$ and $\alpha_{\pi^{\pm}}-\beta_{\pi^{\pm}}$
at backward angles and fit $\xi$ and
$\alpha_{\pi^{\pm}}+\beta_{\pi^{\pm}}$
at forward angles.

\begin{figure}
\epsfysize=8cm
\centerline{\epsffile{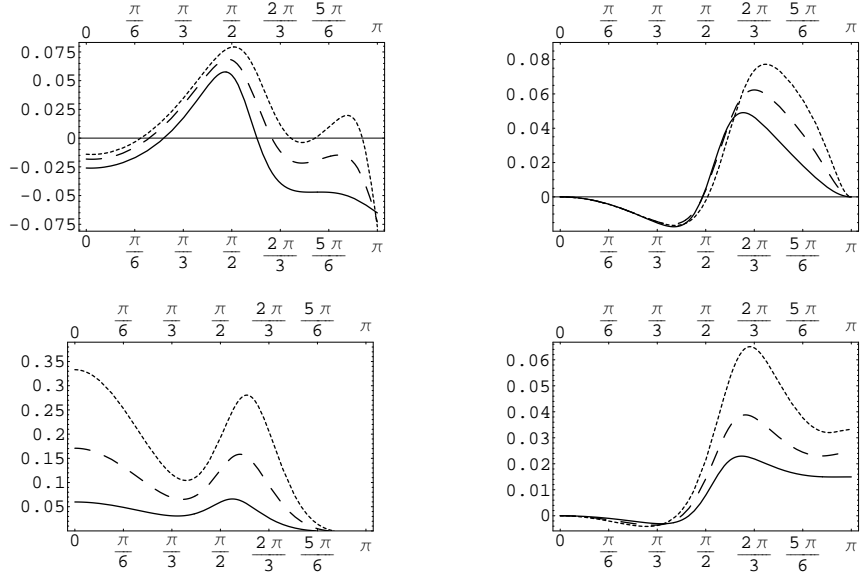}}
\vspace{1cm}
\caption{$R^{\parallel}$'s as functions of $\theta$.
$R_{\xi}^{\parallel}$
(left) and $R_{\eta}^{\parallel}$(right) are in the upper panel.
$R_{+}^{\parallel}$(left) and $R_{-}^{\parallel}$(right) are in the lower
panel.
The solid (dashed, dotted) line is case of $S_{1}$=$4 m_{\pi}^{2}$
($7 m_{\pi}^{2}$,
$10 m_{\pi}^{2}$).
$t$=$-3 m_{\pi}^{2}$ in  all cases}
\label{P}
\end{figure}
\begin{figure}
\epsfysize=8cm
\centerline{\epsffile{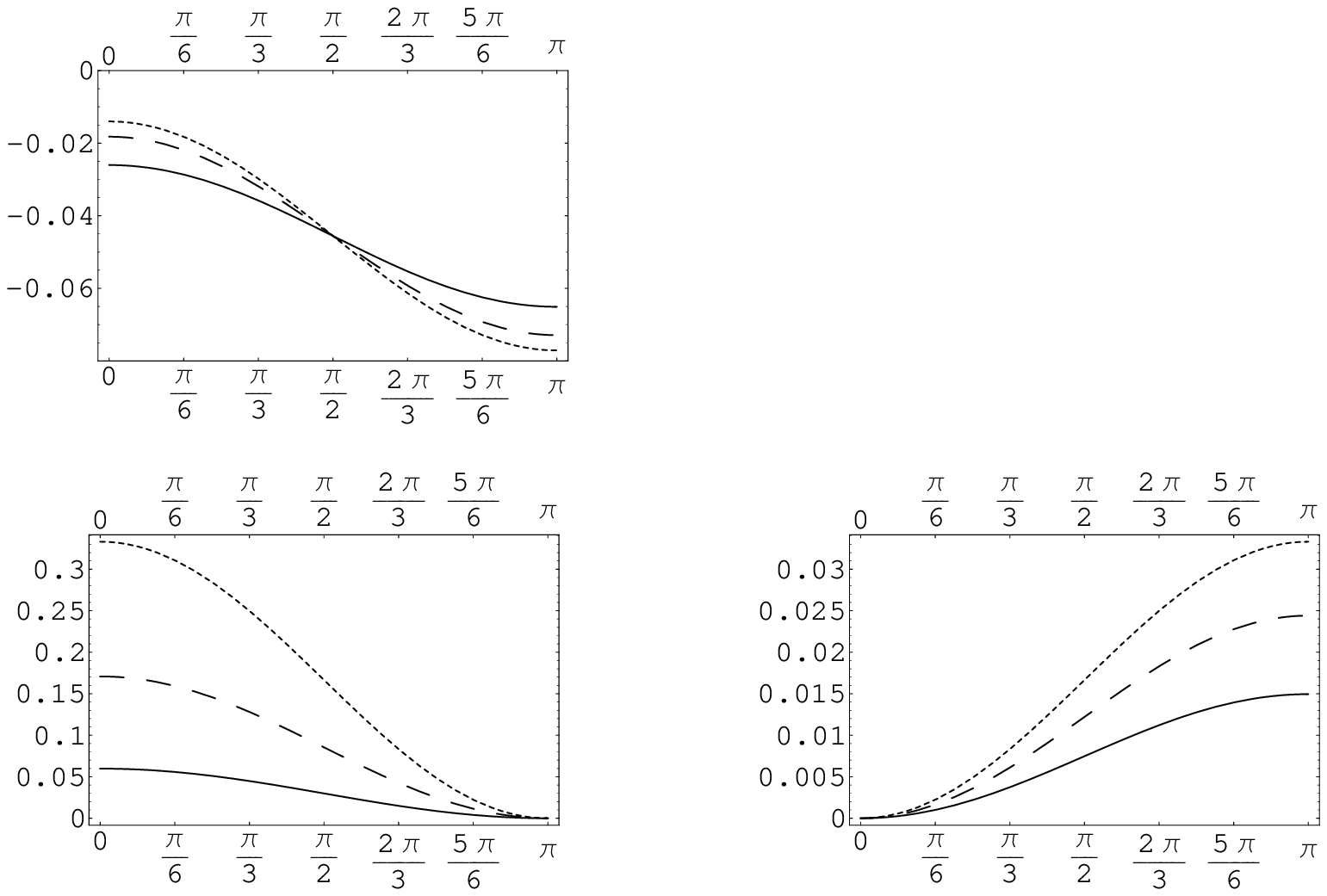}}
\vspace{1cm}
\caption{$R^{\perp}$'s as functions of $\theta$. $R_{\xi}^{\perp}$(left) is
in
the upper panel.
$R_{+}^{\perp}$(left) and $R_{-}^{\perp}$(right) are in the lower panel.
The solid(dashed, dotted) line is case of
$S_{1}$=$4 m_{\pi}^{2}$ ($7 m_{\pi}^{2}$,
$10 m_{\pi}^{2}$).
$t$=$-3 m_{\pi}^{2}$ in all cases.}
\label{N}
\end{figure}
The polarization of the photon has a significant influence on the extraction
of the charged pion polarizabilities.
Consider the Fig.~(\ref{P}), where the polarization vector of the incoming
photon $\vec{\epsilon}_{1}$ is parallel to the scattering plane.
$R_{\eta}^{\parallel}$ is no longer smaller than
$R_{\xi}^{\parallel}$ as in the unpolarized case. One observes the
bumps in $R_{\xi}^{\parallel},\,R_{+}^{\parallel}$ and
$R_{-}^{\parallel}$ between $120^{\circ}\ge\theta\ge 90^{\circ}$.
Those bumps are due to the small values of
$\left(\frac{d^{3}\sigma}{dtdS_{1}d{\Omega}}\right)^{LO}$. Again we
see that it is necessary to take the effects of $\xi$ and $\eta$
into account when trying to extract
$\alpha_{\pi^{\pm}}-\beta_{\pi^{\pm}}$
($\alpha_{\pi^{\pm}}+\beta_{\pi^{\pm}}$ ) at backward (forward)
angles.
\newline\indent
The situation becomes very different when the polarization vector
is perpendicular to the scattering plane. $R_{\eta}^{\perp}$ is
identically zero so the extraction of the charged pion polarizabilities is
simplified. The Fig.(\ref{N}) shows that, in contrast to
$R_{\xi}^{\parallel}$, $R_{\xi}^{\perp}$ decreases with $\theta$
more smoothly. But, $R_{\xi}^{\perp}$ is still comparable to
$R_{-}^{\perp}$ in the backward direction and $R_{+}^{\perp}$ at forward angles.
Therefore one must fit $\xi$ with the charged pion polarizabilities simultaneously.

\section{Applicability of the extrapolation method}
\begin{figure}[t]
\centerline{\epsfxsize 4.0 truein\epsfbox{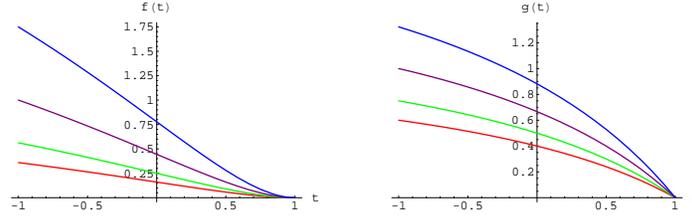}}
\caption{$f(t)=\frac{(t-m_{\pi}^{2})^{2}}{(u-m_{\pi}^{2})^{2}}$ (left panel) and $g(t)=\frac{t-m_{\pi}^{2}}{u-m_{\pi}^{2}}$ (right panel)
as the function of $t$ with $S_{1}=3m_{\pi}^{2}$.
The bottom (red) curve is for $\theta=\pi/3$, the next (green) curve is for $\theta=\pi/2$ and the second(purple) curve is for $\theta=2\pi/3$
and the top (blue) curve is for $\theta=5\pi/6$. The range of $t$ is $-m_{\pi}^{2}\le t\le m_{\pi}^{2}$. The unit of $t$ is $m_{\pi}^{2}$.
}
\label{u-m}
\end{figure}

In this section we discuss the method of extrapolation in \cite{Aiber, DF94}.
This method is to extrapolate the experimental data for
$\gamma+p\rightarrow \gamma+n+\pi^{+}$ obtained near small negative
$t=t_{min}< 0$ to the pion pole $t=m_{\pi}^2$, then
obtains the $\gamma\pi^{+}$ scattering cross
section.
When $|t|$ is small the pion-pole diagrams are expected to be dominant.
Therefore, in \cite{Aiber} only the diagrams with the pole at
$t=m_{\pi}^{2}$, such as the first sixteen diagrams in Fig. 4 , are considered:
\begin{equation}
{\cal A}\simeq{\cal M}(p\rightarrow \pi^{+}+n)\cdot\frac{i}{t-m_{\pi}^{2}}
\cdot{\cal M}(\pi^{+}+\gamma\rightarrow \pi^{+}+\gamma).
\end{equation}
Then one has
\begin{equation}
\frac{d^{3}\sigma_{\gamma N\rightarrow\gamma\pi N}}{dtdS_{1}d{\Omega}}
\propto |{\cal A}~|^{~2}\simeq \frac{|{\cal M}(\pi^{+}+\gamma\rightarrow
\pi^{+}+\gamma)~|^2 \cdot|{\cal M}(p\rightarrow \pi^{+}+n)|^2}
{(t-m_{\pi}^{2})^{2}}.
\end{equation}
Consequently, one obtains the Chew-Low relation
\begin{equation}
\frac{d^{3}\sigma_{\gamma N\rightarrow\gamma\pi N}}
{dtdS_{1}d{\Omega}}=\frac{g^{2}}{4\pi}\frac{S_{1}-m_{\pi}^{2}}{8\pi
M_{p}^{2}E_{\gamma}^{2}}
\frac{(-t)}{(t-m_{\pi}^{2})^{2}}[G_{A}(t)]^{2}
\left(\frac{d\sigma}{d\Omega}\right)_{\gamma \pi\rightarrow \gamma
\pi}+\hat{F}_{B}(S_{1},t,\theta),
\end{equation}
where $\frac{g^{2}}{4\pi}=14.7$ is the pion-nucleon coupling constant
and $\hat{F}_{B}$ represents the contribution of other diagrams
without a pole at $t=m_{\pi}^{2}$. $G_{A}(t)$ is the axial form factor
of the nucleon.
To remove the pole at $t=m_{\pi}^2$, one defines the following quantity:
\begin{eqnarray}
F(t,S_{1},\theta)&\equiv& (t-m_{\pi}^{2})^{2}\cdot (2E_{\gamma}M_{p})^{2}\cdot
\frac{d^{3}\sigma_{\gamma N\rightarrow\gamma\pi N}}
{dtdS_{1}d{\Omega}}\nonumber \\ &\rightarrow&
\frac{-g^{2}}{8\pi^{2}}(S_{1}-m_{\pi}^{2})\cdot t\cdot
[G_{A}(t)]^{2}\left(\frac{d\sigma}{d\Omega}\right)_{\gamma\pi
\rightarrow \gamma\pi}
+(t-m_{\pi}^2)^{2}\cdot F_{B}(t,S_{1},\theta).
\label{Ft}
\end{eqnarray}
Here $F_{B}=\hat{F}_{B}\cdot (2E_{\gamma}M_{p})^{2}$.
The final step is to use measurements of $F(t,S_{1},\theta)$ at
$t_{min}\le 0$ to extrapolate to the pion pole:
\begin{equation}
\lim_{t\rightarrow m_{\pi}^{2}} F(t,S_{1},\theta)\rightarrow
\frac{-g^{2}}{8\pi^{2}}(S_{1}-m_{\pi}^{2})\cdot \,m_{\pi}^{2}\cdot
[G_{A}(t=m_{\pi}^{2})]^{2}\times
\left(\frac{d\sigma}{d\Omega}(S_{1},\theta)\right)
_{\gamma\pi\rightarrow \gamma\pi}.
\label{final}
\end{equation}
Because $F_{B}$ has no double pole at $t=m_{\pi}^{2}$, therefore the second
term of (\ref{Ft}) will decrease rapidly due to the prefactor
$(t-m_{\pi})^{2}$ when the value of $t$ is extrapolated from physical
$t_{min}\le 0$ to $m_{\pi}^{2}$. It is crucial that $F_{B}$ has no singularity
when $0\le t \le m_{\pi}^{2}$.
Furthermore, even $F_{B}$ has no singularity, if
its value becomes large enough to compensate the smallness of the
prefactor $(t-m_{\pi}^{2})^2$ at $t=t_{min}$,
then the validity of Eq.(\ref{final}) will be questionable.
\newline
\indent
Using the amplitudes listed in Eq.(\ref{eq:LO},\ref{eq:NLO})
one can estimate $F_{B}$ and examine whether the extrapolation
is an appropriate procedure.
The method of extrapolation has been successfully used to extract $\pi\pi$
scattering parameters from $\pi N\rightarrow \pi\pi N$.
However,
the extrapolation in the case of radiative charged pion
photoproduction is more complicated because
that there are diagrams which have poles
at $u=m_{\pi}^{2}$ and such diagrams never appear in the case of
$\pi N\rightarrow \pi\pi N$.
Those diagrams have to be included in $\hat{F}_{B}$
since they have no pole at $t=m_{\pi}^{2}$.
Moreover, those diagrams must be included because they are required by gauge
invariance.
According to our results in Eq.(\ref{eq:LO}) and Eq.(\ref{eq:NLO}), at the final $\gamma-\pi$ c.m frame,
$F_{B}(t)$ can be written as
\begin{equation}
\Delta F(t,S_{1},\theta)\equiv (t-m_{\pi}^{2})^{2}\cdot F_{B}(t,S_1,\theta)= F_{B}^{1}(t,S_1,\theta)\frac{(t-m_{\pi}^{2})^{2}}{(u-m_{\pi}^{2})^2}
+ F_{B}^{2}(t,S_1,\theta)\frac{(t-m_{\pi}^{2})}{(u-m_{\pi}^{2})}
+(t-m_{\pi}^{2})^{2}\cdot F^{3}_{B}(t,S_{1},\theta).
\label{FB}
\end{equation}
The value of $\frac{1}{(u-m_{\pi}^{2})^{2}}$
becomes large when the
angle $\theta$ moves toward the backward direction.
Therefore in such situations
the values of $(t-m_{\pi}^2)^{2}\cdot F_{B}$
are not necessarily as small as expected. It can be observed from Fig.(\ref{u-m})
where $f(t)\equiv \frac{(t-m_{\pi}^{2})^{2}}{(u-m_{\pi}^{2})^{2}}$ and $g(t)\equiv
\frac{(t-m_{\pi}^{2})}{(u-m_{\pi}^{2})}$ are plotted as functions
of $t$ and their values increase fast as $\theta$ moves toward the backward direction
in the physical region $t\le 0$.
As a result, the result of the extrapolation derived from Eq. (\ref{final})
will significantly deviate from the correct value
of $\left(\frac{d\sigma}{d\Omega}\right)_{\gamma\pi\rightarrow
\gamma\pi}^{elastic}$, particularly at backward angles, if $F_{B}$ is simply
neglected at $t=t_{min}$.

Hence the applicability of the method of extrapolation
heavily relies on the size of $F^{1}_{B}$ and $F^{2}_{B}$.
One would argue that the amplitude with a pole at $u=m_{\pi}^{2}$ should not hinder the extrapolation as long as one
can calculate the amplitude accurately and include them. However
the LEC $d_{20}$ and $d_{21}+\frac{d_{22}}{2}$ appear in the ${\cal M}_{D}$ and contribute to the cross section through
the diagram (4-19) which owns a pole at $u=m_{\pi}^{2}$. The interference between diagram (4-19) and diagram (1-C) will be comparable with
the interference between diagram (4-1) and diagram (1-A) and causes large deviation of the values of the
charged pion polarizabilities.
Therefore the applicability of the method of extrapolation is severely limited.

Fortunately there is one exception:
when the incoming photon is polarized along the
direction perpendicular to the scattering plane spanned by $\vec{k}$ and
$\vec{r}$, all LO and NLO diagrams with $1/(u-m_{\pi}^{2})$ vanish.
Hence the LO, NLO and NNLO pieces of $F_{B}^{1}$ and $F_{B}^{2}$
all vanish in this polarization condition.
So, the procedure of extrapolation would be applicable
in this particular polarization condition even at the backward angles.
Hence one can apply the method of extrapolation to extract
$\alpha_{\pi^{\pm}}-\beta_{\pi^{\pm}}$ if the incoming photon is polarized
normal to the scattering plane.

\section{Discussion and Conclusion}
\begin{figure}
\epsfysize=1.4cm
\centerline{\epsffile{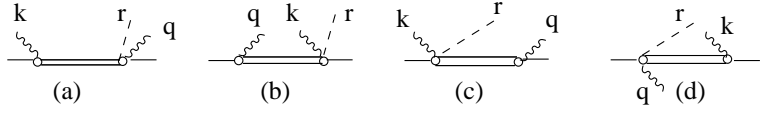}}
\vspace{1cm}
\caption{The LO diagrams of radiative pion electroproduction in ${\cal
M}_{H}$
in the gauge $\epsilon_{1}\cdot v=\epsilon_{2}\cdot v=0$
with the intermediate $\Delta(1232)$ state.}
\label{diagram:delta}
\end{figure}

In this section we need address several important issues.
The first one is the
contribution of nucleon resonances, particularly the contribution of
the $\Delta(1232)$. It is well-known that the $\Delta(1232)$ plays
an important role in the low-energy phenomenology of the nucleon.
It is possible
to include its effect systematically in the extended version of
HBChPT \cite{Delta}. Here we only consider the leading order
contribution of the $\Delta(1232)$ in ${\cal M}_{H}$. The complete
analysis beyond the scope of this article, but the leading order
result in ${\cal M}_{H}$ is very instructive. The LO diagrams with
the $\Delta(1232)$ are given by: (see Fig.~\ref{diagram:delta})
\begin{eqnarray}
{\cal A}_{\Delta}^{a}&=&\frac{-e^{2}g_{\pi\Delta N}b_{1}}{36M_{N}F_{\pi}}
[\tau_{c}-\tau_{3}\delta_{c3}]\frac{1}{\omega_{k}-\Delta}
\left\{-(\vec{\epsilon}_{1}\cdot\vec{\epsilon}_{2}^{~*})
(\vec{\sigma}\cdot \vec{k}~)+(\vec{\epsilon}_{2}^{~*}\cdot \vec{k}~)
(\vec{\sigma}\cdot \vec{\epsilon}_{1})+
2i\,\vec{\epsilon}_{2}^{~*}\cdot\vec{\epsilon}_{1}\times\vec{k}~\right\},
\nonumber
\\{\cal A}_{\Delta}^{b}&=&\frac{-e^{2}g_{\pi\Delta N}b_{1}}{36M_{N}F_{\pi}}
[\tau_{c}-\tau_{3}\delta_{c3}]\frac{1}{\omega_{q}+\Delta}
\left\{-(\vec{\epsilon}_{1}\cdot\vec{\epsilon}_{2}^{~*})
(\vec{\sigma}\cdot \vec{q}~)+(\vec{\epsilon}_{1}\cdot \vec{q}~)
(\vec{\sigma}\cdot \vec{\epsilon}_{2}^{~*})+2i\vec{\epsilon}_{1}\cdot
\vec{\epsilon}_{2}^{~*}\times\vec{q}~\right\}, \nonumber
\\{\cal A}_{\Delta}^{ c}&=&\frac{-e^{2}g_{\pi\Delta N}b_{1}}
{36M_{N}F_{\pi}}[\tau_{c}-\tau_{3}\delta_{c3}]
\frac{1}{\omega_{k}-\omega_{r}-\Delta}\left\{(\vec{\epsilon}_{1}
\cdot\vec{\epsilon}_{2}^{~*})(\vec{\sigma}\cdot \vec{q}~)-
(\vec{\epsilon}_{1}\cdot \vec{q}~)(\vec{\sigma}\cdot
\vec{\epsilon}_{2}^{~*})+2i\,\vec{\epsilon}_{1}\cdot
\vec{\epsilon}_{2}^{~*}\times\vec{q}~\right\}, \nonumber
\\{\cal A}_{\Delta}^{d}&=&\frac{-e^{2}g_{\pi\Delta N}b_{1}}
{36M_{N}F_{\pi}}[\tau_{c}-\tau_{3}\delta_{c3}]\frac{1}
{\omega_{q}+\omega_{r}+\Delta}\left\{(\vec{\epsilon}_{1}
\cdot\vec{\epsilon}_{2}^{~*})(\vec{\sigma}\cdot \vec{k}~)-
(\vec{\epsilon}_{2}^{~*}\cdot \vec{k}~)(\vec{\sigma}
\cdot \vec{\epsilon}_{1})+2i\vec{\epsilon}_{2}^{~*}
\cdot\vec{\epsilon}_{1}\times\vec{k}~\right\},\nonumber \\
\end{eqnarray}
where $\Delta=M_{\Delta}-M_{N}=293 MeV$, $g_{\pi\Delta N}$ is the
$\pi N\Delta$ coupling constant and $b_{1}$ is the coupling constant of
$\gamma N\Delta$ \cite{Delta}.
Large $N_{c}$ QCD gives
\begin{equation}
b_{1}=-\frac{3}{2\sqrt{2}}\kappa_{v},\,\,\,g_{\pi \Delta N}=
\frac{3}{2\sqrt{2}}g_{A}.
\end{equation}
If $\Delta \gg \omega_{k},\omega_{q},\omega_{r}$, then
\begin{equation}
{\cal A}_{\Delta}\simeq\frac{-e^{2}g_{\pi\Delta N}b_{1}}
{18M_{N}\Delta F_{\pi}}[\tau_{c}-\delta_{c3}\tau_{3}]
[(\vec{\epsilon}_{1}\cdot\vec{\epsilon}_{2}^{~*})
(\vec{\sigma}\cdot \vec{k}-\vec{q}~)-(\vec{\sigma}
\cdot \vec{\epsilon_{1}})(\vec{\epsilon}_{2}^{~*}
\cdot \vec{k}~)+(\vec{\sigma}\cdot\vec{\epsilon}_{2}^{~*})
(\vec{\epsilon}_{1}\cdot \vec{q}~)].\label{delta}
\end{equation}
Comparing with Eq.~(\ref{cta}), one finds that it has the same form
as Eq.~(\ref{cta}). If one assumes that $d_{21}+\frac{d_{22}}{2}$ is saturated
by
the $\Delta(1232)$ resonance, then one obtains the following estimate:
\begin{equation}
\xi\equiv (4\pi F_{\pi})^{2}\cdot\left(d_{21}+\frac{d_{22}}{2}\right)
\simeq -\frac{1}{18}\frac{g_{\pi\Delta N}b_{1}}{M_{N}\Delta}
\cdot(4\pi F_{\pi})^{2}\simeq 1.46.
\label{estimate}
\end{equation}
According to the ${\cal O}(p^4)$ CHPT predictions
$\tilde{\alpha}+\tilde{\beta}=0$ and $\tilde{\alpha}-\tilde{\beta}\simeq
1.79$.
Including higher order corrections, the ChPT predictions become
$\tilde{\alpha}+\tilde{\beta}\simeq 0.12$ and
$\tilde{\alpha}-\tilde{\beta}\simeq 1.49$.
According to Eq.(\ref{cros})and Fig. (6-8) the effects due to $\xi$
and the effects due to pion polarizabilities are
about the same magnitude. Hence
it is necessary to take $\xi$ into consideration
when one tries to extract the charged pion polarizabilities from radiative
charged pion photoproduction.
\newline\indent
There is another important issue to be addressed here.
The results of HBChPT for those sub-diagrams such as ${\cal M}_{A}$ and so on
are not unique.
Because each sub-diagram has at least one off-shell external leg,
consequently those
amplitudes are changed if the parameterizations of the pion field and the nucleon
field are changed.
In other words, one can redefine the fields and the results of
${\cal M}_{A}$ to ${\cal M}_{H}$ will be changed.
One might worry about the uniqueness of the result.
As a matter of fact, there is no ambiguity as long as
one uses the same parameterization of the pion and nucleon fields
through the whole calculation.
Because
the physical observables derived from the $S$ matrices with on-shell external legs,
are independent of the choice of
parameterization of pion and nucleon fields.
Note that the explicit forms of the counter terms and the values of the
LECs in the chiral Lagrangian are dependent on the choice of
the parametrization of the field.
Conversely, even if a model is very successful in describing the
experimental data of sub-processes such as
$\gamma+p\rightarrow\gamma+n+\pi^{+}$,
it is not necessarily reliable to use this model to describe the whole process
because of the off-shell ambiguity.
But there is no such an ambiguity in any effective field theory
as long as the physical process is concerned.
That is the main advantage of an
approach based on an effective field theory.
\newline\indent
It is also important to point out that in HBChPT, the convergence of some quantities
such as spin polarizabilities, which are
extracted from some processes such as spin-dependent Compton
scattering off the nucleon, is very poor
\cite{JKO,GHM,KMB}.
It casts doubt on the convergence of the expansion of the amplitude
${\cal A}$ in Eq.(\ref{crosssection}).
However, it has been shown that the convergence of the differential cross
sections for Compton scattering is good \cite{MvGovern}.
The poor convergence of spin polarizabilities is due to the
separation of the nucleon pole and non nucleon-pole contribution of the total amplitude.
Since here we only concern the total amplitude of radiative charged pion photoproduction,
therefore, we should be satisfied without further high-order calculations.
\newline\indent
In conclusion, we find that the the main uncertainty in the extraction of pion
polarizabilities arises from the effect of two combinations of the low
energy
constants in the chiral Lagrangian ${\cal L}_{\pi N}^{(3)}$, $d_{20}$ and
$d_{21}+\frac{d_{22}}{2}$.
Their effects are comparable to the effects of pion polarizabilities on the
cross section for radiative charged pion photoproduction.
Therefore, a measurement with a large coverage of the
scattering angle is required to fit $\xi$, $\eta$ and
$\alpha_{\pi^{\pm}}-\beta_{\pi^{\pm}}$ in the backward direction and to fit
$\xi$ and $\alpha_{\pi^{\pm}}+\beta_{\pi^{\pm}}$ in the forward direction.
We also find the direction of the polarization of the incoming photon
plays important role in the extraction.
Moreover, the typical extrapolation procedure is severely limited due to
diagrams which have pole at $u=m_{\pi}^{2}$ but not at
$t=m_{\pi}^{2}$. However such diagrams will vanish
when the polarization vector
of the incoming photon is perpendicular to the scattering plane. As a result we
suggest that the extrapolation is still applicable in this particular situation.

\section*{acknowledgments}
This work was supported by NSC, Taiwan under Grant No. NSC
96-2112-M-033-003-MY (C.W.K) and
the US Department of Energy under Grant
DE-FG02-97ER41025(B.E.N and K.W.).
We thank Kai Schwenzer for useful suggestions and
careful reading of the manuscript.
\begin{appendix}
\section{Amplitudes of sub-diagram H}
Here we present the amplitudes for the sub-diagrams ${\bf (H)}$.
We use the following notation:
\begin{eqnarray}
\frac{1}{i}\int \frac{d^{d}l}{(2\pi )^{d}}
\frac{1}{(m_{0}^{2}-l^2-i\epsilon)}&=&
\Delta[m_{0}^{2}],\nonumber
\\\frac{1}{i}\int \frac{d^{d}l}{(2\pi )^{d}}\frac{l_{\mu}l_{\nu}}
{(m_{0}^{2}-l^2)^{2}-i\epsilon}&=&g_{\mu\nu}M_{2}[m_{0}^{2}], \nonumber
\\\frac{1}{i}\int \frac{d^{d}l}{(2\pi )^{d}}\frac{1}{(v\cdot l-A-i\epsilon)
(m_{0}^{2}-l^2-i\epsilon)}&=&J_{0}[A,m_{0}^{2}], \nonumber
\\\frac{1}{i}\int \frac{d^{d}l}{(2\pi )^{d}}\frac{l_{\mu}}
{(v\cdot l-A-i\epsilon)(m_{0}^{2}-l^2-i\epsilon)}&=&
v_{\mu}J_{1}[A,m_{0}^{2}], \nonumber
\\\frac{1}{i}\int \frac{d^{d}l}{(2\pi )^{d}}\frac{l_{\mu}l_{\nu}}
{(v\cdot l-A-i\epsilon)(m_{0}^{2}-l^2-i\epsilon)}&=&
g_{\mu\nu}J_{2}[A,m_{0}^{2}]+v_{\mu}v_{\nu}J_{3}[A,m_{0}^2].\nonumber\\
\end{eqnarray}
Using dimensional regularization one obtains
\begin{eqnarray}
\Delta_{0}[m_0^{2}]&=&2m_{0}^{2}
\left(L+\frac{1}{16\pi^2}\ln\frac{m_0}{\mu}\right), \nonumber\\
M_{2}[m_0^2]&=&\frac{1}{d}\left[\Delta_{0}[m_0^2]+
m_0^2\cdot\frac{d\Delta_{0}[m_0^2]}{dm_{0}^2}\right],\nonumber\\
J_{0}[A,m_0^2]&=&-4AL+\frac{A}{8\pi^2}\left(1-2\ln\frac{m_0}{\mu}\right)
-\frac{1}{4\pi^2}\sqrt{m_0^2-A^2}\cdot\cos^{-1}\left(\frac{-A}{m_0}\right),
\nonumber \\
J_{1}[A,m_{0}^{2}]&=&A\cdot J_{0}[A,m_{0}^{2}]+\Delta_{0}[m_0^2],\nonumber
\\
J_{2}[A,m_0^{2}]&=&\frac{1}{d-4}\left[(m_0^2-A^2)J_{0}[A,m_{0}^{2}]-A
\cdot \Delta_{0}[m_0^2]\right],\nonumber\\
J_{3}[A,m_{0}^{2}]&=&A\cdot J_{1}[A,m_0^2]-J_{2}[A,m_0^2],\nonumber\\
L&=&\frac{\mu^{d-4}}{16\pi^2}\left[\frac{1}{d-4}+\frac{1}{2}
\left(\gamma_{E}-1-\ln 4\pi\right)\right], \nonumber\\
\end{eqnarray}
where $\mu$ is the renormalization scale, $\gamma_{E}=0.557215....$ is the
Euler number and $d$ is the space-time dimension.
Also we define $w=(k-q)^{2}$ and
$J^{(m)}_{0}\equiv\frac{\partial^{m}J_{0}[A,m_{0}^{2}]}
{\partial (m_{0}^{2})^{m}}$.
The results are
\begin{eqnarray}
&&{\cal M}_{H1}=\frac{-\sqrt{2}e^{2}g_{A}^{2}}
{2F_{\pi}^{3}}\int^{1}_{0}dy\int^{1-y}_{0}dx\nonumber
\\&&(\vec{\epsilon_{1}}\cdot\vec{\epsilon}_{2}^{~*})
(\vec{\sigma}\cdot \vec{k})(x+y)\left[M_{2}^{(1)}[m_{\pi}^{2}+y(y-1)w+xyw]
+2\omega_{r}J_{2}^{(2)}[x\omega_{k}+y\omega_{r},m_{\pi}^{2}+y(y-1)w+xyw]\right]
\nonumber
\\&+&(\vec{\epsilon}_{1}\cdot\vec{\epsilon}_{2}^{~*})
(\vec{\sigma}\cdot \vec{q})(-y)\left[M_{2}^{(1)}[m_{\pi}^{2}+y(y-1)w+xyw]+2
\omega_{r}J_{2}^{(2)}[x\omega_{k}+y\omega_{r},m_{\pi}^{2}+y(y-1)w+xyw]\right]
\nonumber
\\&+&(\vec{\sigma}\cdot \vec{\epsilon}_{1})(\vec{\epsilon}_{2}^{~*}
\cdot \vec{k})(x+y-\frac{1}{2})\left[M_{2}^{(1)}[m_{\pi}^{2}+y(y-1)w+xyw]+2
\omega_{r}J_{2}^{(2)}[x\omega_{k}+y\omega_{r},m_{\pi}^{2}+y(y-1)w+xyw]\right]
\nonumber
\\&+&(\vec{\sigma}\cdot\vec{\epsilon}_{2}^{~*})(\vec{\epsilon}_{1}
\cdot \vec{q})(x+y-1)\left[M_{2}^{(1)}[m_{\pi}^{2}+y(y-1)w+xyw]+2
\omega_{r}J_{2}^{(2)}[x\omega_{k}+y\omega_{r},m_{\pi}^{2}+y(y-1)w+xyw]\right]
\nonumber
\\&+&(\vec{\epsilon_{1}}\cdot \vec{q})(\vec{\epsilon}_{2}^{~*}\cdot \vec{k})
(\vec{\sigma}\cdot\vec{k})[y(x+y)(x+y-1)]\left[\Delta_{0}^{(2)}[m_{\pi}^{2}+
y(y-1)w+xyw]+2\omega_{r}J_{0}^{(2)}[x\omega_{k}+y\omega_{r},m_{\pi}^{2}+
y(y-1)w+xyw]\right] \nonumber
\\&-&(\vec{\epsilon}_{1}\cdot \vec{q})(\vec{\epsilon}_{2}^{~*}\cdot \vec{k})
(\vec{\sigma}\cdot\vec{q}) y^{2}(x+y-1)\left[\Delta_{0}^{(2)}
[m_{\pi}^{2}+y(y-1)w+xyw]+2\omega_{r}J_{0}^{(2)}[x\omega_{k}+
y\omega_{r},m_{\pi}^{2}+y(y-1)w+xyw]\right]\nonumber
\\\end{eqnarray}
\begin{eqnarray}
&&{\cal M}_{H2}=\frac{-\sqrt{2}e^{2}g_{A}^{2}}{2F_{\pi}^{3}}\int^{1}_{0}dy
\int^{1-y}_{0}dx\nonumber
\\&&(\vec{\epsilon}_{1}\cdot\vec{\epsilon}_{2}^{~*})(\vec{\sigma}\cdot
\vec{q})
(x+y)\left[M_{2}^{(1)}[m_{\pi}^{2}+y(y-1)w+xyw]-2\omega_{r}J_{2}^{(2)}
[x\omega_{q}-y\omega_{r},m_{\pi}^{2}+y(y-1)w+xyw]\right]\nonumber
\\&+&(\vec{\epsilon}_{1}\cdot\vec{\epsilon}_{2}^{~*})
(\vec{\sigma}\cdot \vec{k})(-y)\left[M_{2}^{(1)}[m_{\pi}^{2}+y(y-1)w+xyw]-2
\omega_{r}J_{2}^{(2)}[x\omega_{q}-y\omega_{r},m_{\pi}^{2}+y(y-1)w+xyw]\right]
\nonumber
\\&+&(\vec{\sigma}\cdot \vec{\epsilon}_{2}^{~*})(\vec{\epsilon}_{1}\cdot
\vec{q})(x+y-\frac{1}{2})\left[M_{2}^{(1)}
[m_{\pi}^{2}+y(y-1)w+xyw]-2\omega_{r}J_{2}^{(2)}
[x\omega_{q}-y\omega_{r},m_{\pi}^{2}+y(y-1)w+xyw]\right]\nonumber
\\&+&(\vec{\sigma}\cdot\vec{\epsilon}_{1})(\vec{\epsilon}_{2}^{~*}\cdot
\vec{k})(x+y-1)\left[M_{2}^{(1)}[m_{\pi}^{2}+y(y-1)w+xyw]-2
\omega_{r}J_{2}^{(2)}[x\omega_{q}-y\omega_{r},m_{\pi}^{2}+y(y-1)w+xyw]\right]
\nonumber
\\&+&(\vec{\epsilon}_{2}^{~*}\cdot \vec{k})(\vec{\epsilon}_{1}\cdot \vec{q})
(\vec{\sigma}\cdot \vec{q})[y(x+y)(x+y-1)]
\left[\Delta_{0}^{(2)}[m_{\pi}^{2}+y(y-1)w+xyw]-2\omega_{r}J_{0}^{(2)}
[x\omega_{q}-y\omega_{r},m_{\pi}^{2}+y(y-1)w+xyw]\right] \nonumber
\\&-&(\vec{\epsilon}_{2}^{~*}\cdot \vec{k})(\vec{\epsilon}_{1}\cdot \vec{q})
(\vec{\sigma}\cdot \vec{k}) y^{2}(x+y-1)\left[\Delta_{0}^{(2)}[m_{\pi}^{2}+y
(y-1)w+xyw]-2\omega_{r}J_{0}^{(2)}[x\omega_{q}-y\omega_{r},m_{\pi}^{2}+y(y-1)
w+xyw]\right]\nonumber \\
\end{eqnarray}
\begin{eqnarray}{\cal M}_{H3}&=&\frac{\sqrt{2}e^{2}g_{A}^{2}}{4F_{\pi}^{3}}
(\vec{\sigma}\cdot \vec{k}-\vec{q})(\vec{\epsilon}_{1}\cdot
\vec{\epsilon}_{2}^{~*})\int^{1}_{0}dx x\cdot\{ \Delta_{0}^{(1)}
[m_{\pi}^{2}+x(x-1)w]+2\omega_{r}J_{0}^{(1)}[x\omega_{r},m_{\pi}^{2}+x(x-1)w]\}
.
\nonumber \\
\end{eqnarray}
\begin{eqnarray}
{\cal M}_{H4}&=&-\frac{\sqrt{2}e^{2}g_{A}^{2}}{4F_{\pi}^{3}}(\vec{\sigma}
\cdot \vec{k}-\vec{q})(\vec{\epsilon}_{1}\cdot\vec{\epsilon}_{2}^{~*})
\int^{1}_{0}dx (1-x)\cdot\{
\Delta_{0}^{(1)}[m_{\pi}^{2}+x(x-1)w]+2\omega_{r}
J_{0}^{(1)}[(x-1)\omega_{r},m_{\pi}^{2}+x(x-1)w]\}.
\nonumber \\
\end{eqnarray}
\begin{equation}
{\cal M}_{H5}={\cal M}_{H6}=0.
\end{equation}
\begin{eqnarray}
{\cal M}_{H7}&=&-\frac{\sqrt{2}e^{2}g_{A}^{2}}{F_{\pi}^{3}}
(\vec{\sigma}\cdot 3\vec{k}-3\vec{q}-2\vec{r})\{(\vec{\epsilon}_{1}\cdot
\vec{\epsilon}_{2}^{~*})\int^{1}_{0}dy\int^{1-x}_{0}dy (-1)\cdot M_{2}^{(1)}
[m_{\pi}^{2}+y(y-1)w+xyw]\nonumber \\&+&(\vec{\epsilon}_{1}\cdot
\vec{q})(\vec{\epsilon}_{2}^{~*}\cdot \vec{k})\int^{1}_{0}dy
\int^{1-x}_{0}dy~y(1-x-y)\Delta_{0}^{(2)}[m_{\pi}^{2}+y(y-1)w+xyw]\}.
\end{eqnarray}
\begin{eqnarray}
{\cal M}_{H8}&=&\frac{\sqrt{2}e^{2}g_{A}^{2}}{2F_{\pi}^{3}}
(\vec{\sigma}\cdot 3\vec{k}-3\vec{q}-2\vec{r})(\vec{\epsilon}_{1}\cdot
\vec{\epsilon}_{2}^{~*})\int^{1}_{0}dx(-1)\cdot\{ \Delta^{(1)}_{0}
[m_{\pi}^{2}+x(x-1)w]\}.
\end{eqnarray}
\begin{equation}
{\cal M}_{H9}=\frac{\sqrt{2}e^{2}}{F_{\pi}}(d_{21}+\frac{d_{22}}{2})
[(\vec{\epsilon_{1}}\cdot\vec{\epsilon}_{2}^{~*})
(\vec{\sigma}\cdot \vec{k}-\vec{q}~)-(\vec{\sigma}\cdot \vec{\epsilon}_{1})
(\vec{\epsilon}_{2}^{*}\cdot \vec{k}~)+(\vec{\sigma}\cdot
\vec{\epsilon}_{2}^{~*})(\vec{\epsilon_{1}}\cdot \vec{q})]
\end{equation}
Here we do not explicitly display the results of cross diagrams which can
be obtained by exchanging
$\vec{k}\longleftrightarrow -\vec{q},\,\,\vec{\epsilon}_{1}
\longleftrightarrow \vec{\epsilon}_{2}^{~*}.$
\end{appendix}

\end{document}